\documentclass[paper]{JHEP3}
\usepackage{booktabs}
\usepackage{psfig}
\usepackage{epsfig,bm,amsmath}
\usepackage{enumerate}
\usepackage{cite}
\preprint{Cavendish--HEP--08/09}
{\title{Applying the {\tt POWHEG} method to top pair production and decays at the ILC}}
\author{Oluseyi Latunde-Dada\\
  Cavendish Laboratory, University of Cambridge,\\
  JJ Thomson Avenue, Cambridge CB3 0HE, U.K.\\
  E-mail: \email{seyi@hep.phy.cam.ac.uk}}

\abstract{We study the effects of gluon radiation in top pair production and their decays for
  $e^+e^-$ annihilation at the ILC. To achieve this we apply the {\tt POWHEG} method and
  interface our results to the Monte Carlo event generator \textsf{Herwig++}. We consider a center-of-mass energy of $\sqrt{s}=500$
  GeV and compare decay correlations and bottom quark and anti-quark distributions before
  hadronization.}

\begin{document}
\section{Introduction}
High energy polarized $e^+e^-$ colliders will be essential instruments in the search for
the fundamental constituents of matter and their interactions. One such collider being
designed is the International Linear Collider (ILC) which is expected to run at
centre-of-mass energies $\geq 500$ GeV. At these energies there will be a significant
proportion of top quark pairs produced from the annihilation process so the ILC provides
an impressive tool to carry out detailed studies of top quark physics. The top quark radiates gluons both in its production phase and its
decay phase, thus it is useful to see how the leading order experimental analysis will be
affected by these QCD corrections.  
There have been numerous studies of top quark production and their decays both at
leading order and next-to leading order in QCD. Details can be found in
\cite{Jezabek:1988iv,Jezabek:1988ja,Czarnecki:1990pe,Bernreuther:1991hy,Schmidt:1995mr,Kodaira:1998gt,Macesanu:1998ts,Nasuno:1999fj}
and many more besides.
In this paper we consider the process at
next-to-leading order in the production and semi-leptonic decays of the top pairs using the {\tt POWHEG} method \cite{Nason:2004rx,Frixione:2007vw} in conjunction with
the Monte Carlo event generator \textsf{Herwig++}\cite{Bahr:2008pv}. The {\tt POWHEG} method has been successfully applied to $Z$ pair
production \cite{Nason:2006hfa}, heavy flavour production \cite{Frixione:2007nw}, $e^+e^-$ annihilation into
hadrons and Drell-Yan vector boson production \cite{Alioli:2008gx,Hamilton:2008pd}.
We work in the narrow width approximation and hence do not include
interference between the production and decay emissions which are negligible in this limit
\cite{Khoze:1992rq,Orr:1993kd}. We also take account of the beam polarization and spin correlations of
the top pairs. Finally, we present plots of some relevant distributions.
\section{Hardest emission generation: Production}
\label{POWHEG}
The order-$\alpha_{s}$ differential cross section for the process $e^+e^- \rightarrow V \rightarrow
t\bar{t}g$ where $V$ represents a vector current, can be written as 
\begin{equation}
\centering
\label{eq:WV}
R(x,y)=\sigma_{V}W_{V}(x,y)=\frac{\sigma_{V}}{v}\frac{2{\alpha}_s}{3{\pi}}\left[\frac{{(x+2
      \rho)}^2+{(y+2 \rho)}^2+\zeta_{V}}{(1+2 \rho)(1-x)(1-y)}-\frac{2 \rho}{{(1-x)}^2}-\frac{2 \rho}{{(1-y)}^2}\right]
\end{equation}
where $\sigma_{V}$ is the Born cross section for heavy quark production, $W_{V}=R/\sigma_{V}$,
$\rho={m_{\rm t}}^2/s$ where $m_{\rm t}$ is the mass of the top quark and $s$ is the square of the center of mass energy,
$\zeta_{V}=-8 \rho(1+2 \rho)$, $v=\sqrt{1-4\rho}$ and $x,y$ are the energy fractions of $t$ and
$\bar{t}$ respectively.

In the case of the axial current contribution $e^+e^- \rightarrow A \rightarrow
t\bar{t}g$, we have
\begin{equation}
\centering
\label{eq:WA}
R(x,y)=\sigma_{A}W_{A}(x,y)=\frac{\sigma_{A}}{v}\frac{2{\alpha}_s}{3{\pi}}\left[\frac{{(x+2
      \rho)}^2+{(y+2 \rho)}^2+\zeta_{A}}{(1-4 \rho)(1-x)(1-y)}-\frac{2 \rho}{{(1-x)}^2}-\frac{2 \rho}{{(1-y)}^2}\right]
\end{equation}
where $\sigma_{A}$ is the Born cross section for heavy quark production by the axial
current, $W_{A}=R/\sigma_{A}$
and $\zeta_{A}=2 \rho[(3+x_{g})^{2}-19+4 \rho]$ where $x_g=2-x-y$.

Because of the top mass, the
phase space for gluon emission is reduced and the collinear divergences present in the
massless quark cross-section are regularized here. However, the infra-red divergence
as the gluon momentum goes to zero is still present. We can write down the cross-section
for the hardest emission as
\begin{equation}
\centering
\label{eq:sig}
d \sigma=\sum \bar {B}(v) d \Phi_{v}\left[\Delta^{(NLO)}_{R}(0)+\Delta^{(NLO)}_{R}(p_T)\frac{R(v,r)}{B(v)}d \Phi_{r}\right]
\end{equation} 
where $B(v)$ is the Born cross section and $v$ represents the Born variables, $r$ represents the
radiation variables and $d \Phi_{v}$ and $d
\Phi_{r}$ are the Born and real emission phase spaces respectively.
 $\Delta_{R}^{NLO}(p_T)$ is the modified Sudakov form factor for the hardest emission with
transverse momentum $p_T$, as indicated by the Heaviside function in the exponent of
(\ref{eq:dnlo}),
\begin{equation}
\centering
\label{eq:dnlo}
\Delta_{R}^{NLO}(p_T)=\exp \left[-\int d\Phi{r}\frac{R(v,r)}{B(v)}\Theta(k_T(v,r)-p_T)\right]\;.
\end{equation}
Furthermore, 
\begin{equation}
\centering
\label{eq:B}
\bar{B}(v)=B(v)+V(v)+\int (R(v,r)-C(v,r))d \Phi_{r}\;.
\end{equation}
$\bar{B}(v)$ is the sum of the Born, $B(v)$, virtual, $V(v)$ and real, $R(v,r)$ terms, (with some
counter-terms, $C(v,r)$).  It overcomes the problem of negative weights
since in the region where $\bar{B}(v)$ is negative, the NLO negative terms must have overcome the Born
term and hence perturbation theory must have failed. It is used to generate the variables of the Born subprocess to which the
real-emission contributions factorize in the collinear limit. 

Now explicitly for $e^+e^- \rightarrow
t\bar{t}g$,
\begin{equation}
\centering
\label{eq:delta}
\Delta_{R}^{NLO}(p_T)=\exp \left[-\int dx\,dy
 W(x,y)\Theta(k_T(x,y))-p_T)\right]
\end{equation}
where
\begin{equation}
\centering
\label{eq:kT}
k_T(x,y)=\sqrt{s\frac{(1-x)(1-y)(x+y-1)-\rho (2-x-y)^2}{\max(x,y)^2-4 \rho}}
\end{equation}
is the transverse momentum of the hardest emitted gluon relative to the splitting axis, as
illustrated in Figure \ref{fig:kt} below.
\begin{figure}[!ht]
\begin{center}
\psfig{figure=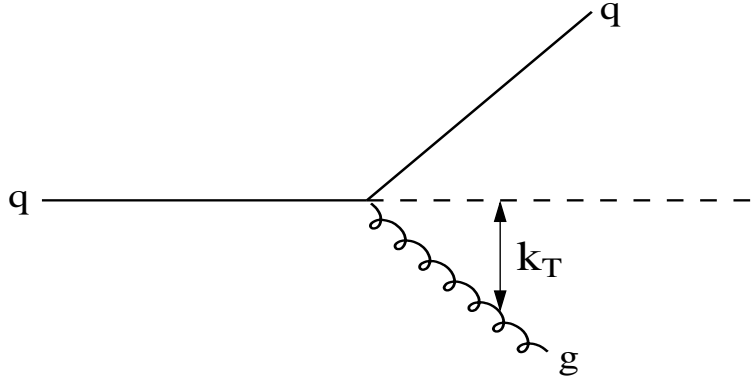,%
width=4in,height=2in,angle=0}
\end{center}
\caption{Transverse momentum, $k_T$.}
\label{fig:kt}
\end{figure}

\subsection{Generation of radiation variables, $x$ and $y$}
\label{sec:xy}
The radiation variables, $x$ and $y$ are to be generated according to the probability
distribution
\begin{equation}
\centering
\label{eq:delW}
\Delta^{W}(k_T)W(x,y)dx\,dy
\end{equation}
where for $e^+e^-$ annihilation via a vector and axial current, $W(x,y)$ and $\Delta^{W}(k_T)$ are
defined in (\ref{eq:WV}),(\ref{eq:WA}) and (\ref{eq:delta}).
 
In the region where $x>y$, let us define the dimensionless variable, $\kappa$ as
\begin{equation}
\centering
\label{eq:kappa}
\kappa=\frac{k_{T}^{2}}{s}=\frac{(1-x)(1-y)(x+y-1)-\rho (2-x-y)^2}{x^2-4 \rho}\;.
\end{equation}

There are two solutions for $y$ for each value of $x$ and $\kappa$.

\begin{equation}
\centering
y_{1,2}=\frac{x^2-3x-2 \rho x+2+4 \rho\pm \sqrt{(x^2-4 \rho)(4 \kappa(x-1- \rho)+(x-1)^2)}}{2(1+\rho-x)}\;.
\label{eq:y1y2}
\end{equation}

Exchanging the $y$ variable for $\kappa$, we find
\begin{eqnarray}
\centering
\label{eq:intt}
\int W(x,y) {\Theta}(k_T(x,y)-p_T)dxdy &=& \int^{x_{\rm max}}_{x_{\rm min}} dx
\int^{\kappa_{\rm max}}_{\kappa}
  d \kappa\frac{2{\alpha}_s(\kappa s)}{3{\pi}}\frac{dy}{d \kappa}W(x,y_{1,2})\nonumber \\
&=&\int dx \int 
  d \kappa\frac{2{\alpha}_s(\kappa
    s)}{3{\pi}} \frac{\sqrt{x^2-4 \rho}}{\sqrt{(4 \kappa(x-1- \rho)-(x-1)^2)}}W(x,y_{1,2})\nonumber \\
\end{eqnarray}
Now the integrand, $W^{'}=\frac{\sqrt{x^2-4 \rho}}{\sqrt{(4 \kappa(x-1- \rho)-(x-1)^2)}}W(x,y_{1,2})$ in (\ref{eq:intt}) yields a complicated integral so we look for an upper
bound on $W^{'}$ which we denote as $V^{'}=\frac{dy}{d \kappa}V(x,y_{1,2})$ to simplify the integration. We then generate the radiation variables as
outlined below:
\begin{enumerate}
\item
Set $p_{\rm max}={k_{T}}_{\rm max}$.\\
\item
For a random number, $n$ between $0$ and $1$, solve the equation below for $p_T$ 
\begin{equation}
\centering
\label{eq:n}
n=\frac{{\Delta^{V}(p_T)}}{\Delta^{V}(p_{\rm max})}\,
\end{equation}
where $\Delta^{V}(p_T)=\exp \left[-\int dx\,d\kappa
V^{'}(x,\kappa)\right]$ 
\item
Generate the variables $x$ and $y$ according to the distribution
\begin{equation}
\centering
\label{eq:Udel}
V(x,y){\delta}(k_T(x,y)-p_T)\;.
\end{equation}
\item
Accept the generated value of $p_T$ with probability $W/V$. If the event is rejected set
$p_{\rm max}=p_T$ and go to 2).
\end{enumerate}

 Using our knowledge of the form of the integrand
in the massless quark case \cite{LatundeDada:2006gx}, we guess that $V^{'}$ should take the form,
\begin{equation}
\centering
\label{eq:V}
V^{'}(x,\kappa)=N_{\kappa}\frac{2{\alpha}_s(\kappa
    s)}{3{\pi}} \frac{4}{(1-x+\gamma(\kappa,x))\gamma(\kappa,x)}
\end{equation}
 where 
\begin{equation}
\centering
\label{eq:rho}
\gamma(\kappa,x)=\sqrt{(1-x)(1-2 \kappa-2 \sqrt{ \kappa^2+\rho \kappa})}\;.
\end{equation}
and $N_{\kappa}$ is a normalisation factor which depends on $\kappa$, which has to be tuned
to ensure that $V^{'}$ is an upper bound of $W^{'}$. Both $V^{'}$ and $W^{'}$ have the same
divergent behaviour at $x_{\rm max}=1-2 \kappa-2 \sqrt{ \kappa^2+\rho\kappa}$.

Table \ref{tab:kap} shows the $N_{\kappa}$ values for different ranges of
$\kappa$ and the two solutions for $y$ used for both axial and vector currents. The lower limit on $\kappa$ was set by choosing $k_T=\Lambda_{\rm
  QCD}=0.2$ GeV thus setting a lower bound on the transverse momentum.
\begin{table}[!ht]
\centering
\begin{tabular}{llll} \hline
Range of $\kappa$ & $N_{\kappa}(y_{1})$&$N_{\kappa}(y_{2})$\\
\hline
\hline                      
$0.024-0.03$  & $0.4$ & $0.4$ \\ 
$0.015-0.024$   & $0.7$& $0.7$ \\ 
$0.005-0.015$ & $1.2$ & $1.1$   \\ 
$0.0005-0.005$ & $4.0$ & $2.6$    \\ 
$0.0001-0.0005$ & $9.0$ &$6.0$   \\ 
$0.00005-0.0001$ & $13.0$& $7.0$\\ 
$0.00003-0.00005$ &$17.0$ & $10.0$   \\ 
$0.00000016-0.00003$ & $45.0$ &$35.0$    \\ 
\hline
\hline
\\
\end{tabular}
\caption{$N_{\kappa}$ for different values of $\kappa$ for both axial and vector currents}
\label{tab:kap}
\end{table}

\begin{figure}[!ht]
\begin{center}
\psfig{figure=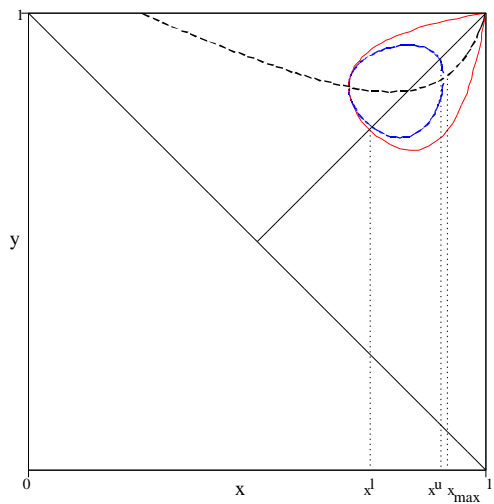,%
width=3in,height=3in,angle=0}
\end{center}
\caption{Phase space and $y$ solutions for $\kappa=0.01$ in the region $x>y$.}
\label{fig:lt}
\end{figure}

\begin{figure}[!ht]
\begin{center}
\psfig{figure=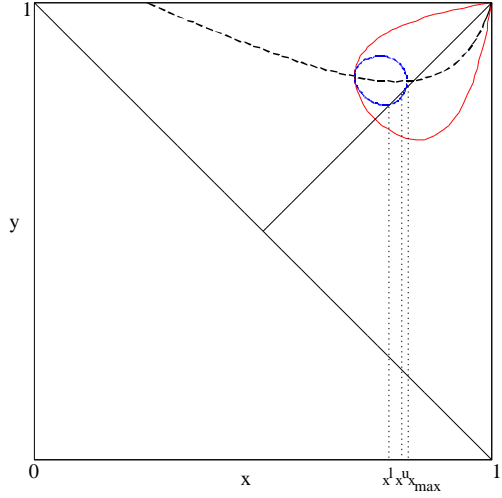,%
width=3in,height=3in,angle=0}
\end{center}
\caption{Phase space and $y$ solutions for $\kappa=0.028$ in the region $x>y$.}
\label{fig:gt}
\end{figure}

We now consider the specific case where $m_{\rm t}=175$ GeV and
$\sqrt{s}=500$ GeV i.e. $\rho=0.1225$. For $\kappa \le 0.024$, there are two $y$ solutions in the region of phase space where
$x>y$. This is illustrated in Figure \ref{fig:lt} below for $\kappa = 0.01$. 
The red line denotes the phase space for gluon emissions. The two solutions lie on either side of
the dashed line 
\begin{equation}
y=\frac{(2-x)(1-x+2\rho)}{2(1-x+ \rho)}
\end{equation}
 and are
equal when 
\begin{equation}
x=x_{\rm max}=1-2 \kappa-2 \sqrt{ \kappa^2+\rho\kappa}
\end{equation}
 which lies on the dotted line. At $\kappa=0.024$, the branches
meet along the line $y=x$ and there is only
one solution for $y$ in the region (the lower branch). So for $\kappa>0.024$, only one $y$
solution exists for $x > y$. This is
illustrated in Figure \ref{fig:gt} for $\kappa=0.028$ . In addition there are no $y$
solutions for $\kappa>0.03$ for $x > y$ . 
Also note that for $x<x^{u}$, there is only one solution for $y$.

In the region where there are two solutions, the integral (with $V^{'}$ in place of $W^{'}$) in (\ref{eq:intt}) is performed along
both branches independently and summed. For the upper branch, $x$ runs from $x^{u}$ to
$x_{\rm max}=1-2 \kappa-2 \sqrt{ \kappa^2+\rho\kappa}$  while for the lower branch, $x$ runs from $x^{l}$ to $x_{\rm max}$ where
if we define

\begin{eqnarray}
\label{eq:x}
x_a&=&39\kappa-1+12\rho-168\rho\kappa-48{\rho}^2+\kappa^3+15\kappa^2+12\rho{\kappa}^2+48{\rho}^2{\kappa}+64{\rho}^3\,,\nonumber \\
x_b&=&6(-33\kappa^2+3\kappa-3\kappa^3+288{\rho}^2\kappa-768{\rho}^3\kappa-48\rho\kappa+168\rho{\kappa}^3-204{\rho}^2{\kappa}^2+300\rho{\kappa}^2
\nonumber \\
&+&12\rho{\kappa}^4+768{\rho}^4\kappa+144{\rho}^2{\kappa}^3+486{\rho}^3{\kappa}^2)^{\frac{1}{2}}\,,\nonumber \\
x_c&=&\sqrt{{x_a}^2+{x_b}^2}\,,\nonumber \\
x_d&=&\tan^{-1}\left(\frac{x_b}{x_a}\right)\,,\nonumber \\
x_e&=&-\frac{1}{12}x_c^{\frac{1}{3}}\cos\left(\frac{x_d}{3}\right)\,,\nonumber \\
x_f&=&\frac{(-1-\kappa^2-10\kappa+8\rho-2\rho\kappa-16{\rho}^2)\cos\left(\frac{x_d}{3}\right)}{12x_c^{\frac{1}{3}}}\,,\nonumber \\
x_g&=&\frac{\kappa+5+4\rho}{6}\,, \nonumber \\
x_h&=&\frac{\sqrt{3}}{12}\sin\left(\frac{x_d}{3}\right)\left(x_c^{\frac{1}{3}}+\frac{1+\kappa^2+10\kappa-8\rho+2\rho\kappa+16{\rho}^2}{x_c^{\frac{1}{3}}}\right)\,,
\end{eqnarray}
we can write $x^{u}$ and $x^{l}$ as
\begin{eqnarray}
\centering
\label{eq:xuxl}
x^{u}&=&x_e+x_f+x_g+x_h\,, \nonumber \\
x^{l}&=&x_e+x_f+x_g-x_h \,;
\end{eqnarray}
In the region where there is only one solution for $y$, $x$ runs from $x^{l}$ to $x^{u}$.\\
The $\kappa$ integration can then be performed numerically. Having performed the
integration, values for $\kappa$ and hence $k_T$ are then generated according to steps 1 and 2 in
Section \ref{sec:xy}. The variables $x$ and $y$ are then to be
distributed according to $W^{'}(x,y){\delta}(k_T(x,y)-p_T)$. This is the subject of the next
section.
\subsection{Distributing $x$ and $y$ according to $W(x,y)$}
\label{sec:Vxy}
To generate $x$ and $y$ values with a distribution proportional to
$V(x,y){\delta}(k_T(x,y)-p_T)$, where from (\ref {eq:intt}), $V(x,y)$ is the $\frac{V^{'}}{dy/d \kappa}$, we can use the $\delta$ function to eliminate the $y$
variable by computing
\begin{equation}
\centering
\label{eq:D}
D(x)=\int dy \delta(k_T-p_T)V(x,y)= \left.\frac{V(x,y)}{\frac{\partial k_T}{\partial y}}\right|_{y=\bar{y}}
\end{equation}
where $\bar{y}$ is such that $k_T(x,\bar{y})=p_T$. Note that $\frac{\partial k_T}{\partial
  y}$ is the same for both $y$ solutions. We then generate $x$ values with a
probability distribution proportional to $D$ with hit-and-miss techniques as described below. All events
generated have uniform weights. 
\begin{enumerate}
\item
Randomly sample $x$, $N_x$ times (we used $N_x=10^5$) in the range $[x_{\rm min}:x_{\rm max}]$ for the selected value of $\kappa$.\\
\item
For each value of $x$, evaluate $\bar{D}=D(x,y_1)+D(x,y_2)$ if there are 2 solutions for the selected $\kappa$ and
$\bar{D}=D(x,y_2)$ if there is only one solution. Also, if $\kappa<0.024$ and $x<x^{u}$ (see
Figure \ref{fig:lt}), there is only one $y$ solution so evaluate $\bar{D}=D(x,y_2)$.\\
\item 
Find the maximum value $\bar{D}_{\rm max}$ of $\bar{D}$ for the selected value of $\kappa$.\\
\item
Next, select a value for $x$ in the allowed range and evaluate $\bar{D}$.\\
\item
If  $\bar{D}> r\bar{D}_{\rm max}$ (where $r$ is a random number between $0$ and $1$),
accept the event, otherwise go to 4.)  and
generate a new value for $x$.\\
\item
If for the chosen value of $x$, there are two solutions for $y$, select a value for $y$ in
the ratio $D(x,y_1):D(x,y_2)$.
\item
Compare $V^{'}(x,y)$ with the true integrand, $W^{'}(x,y)$. If the event fails this veto, set
$\kappa_{\rm max}=\kappa$ and
  regenerate a new $\kappa$ value as discussed in Section \ref{sec:xy}.\\
\end{enumerate}  
NB: For the region $y>x$, exchange $x$ and $y$ in the above discussion.
In this way, the smooth phase space distribution in Figure \ref{fig:pa} below was obtained
for the
hardest emission events for an axial current. The plot show 2,500 of these events.
\begin{figure}[!ht]
\vspace{2cm}
\hspace{0.5cm}
\psfig{figure=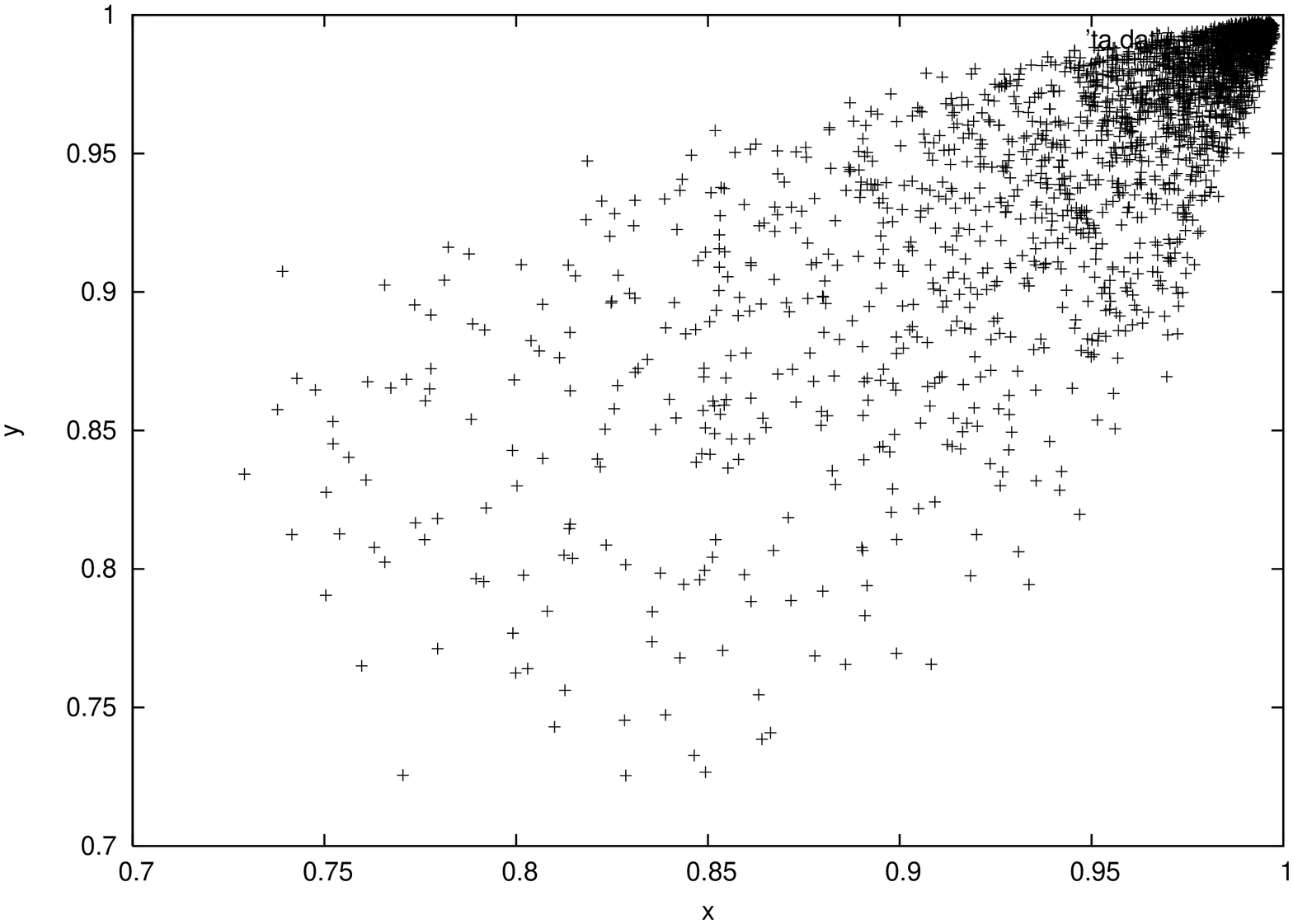,%
width=2.5in,height=3in,angle=270}
\psfig{figure=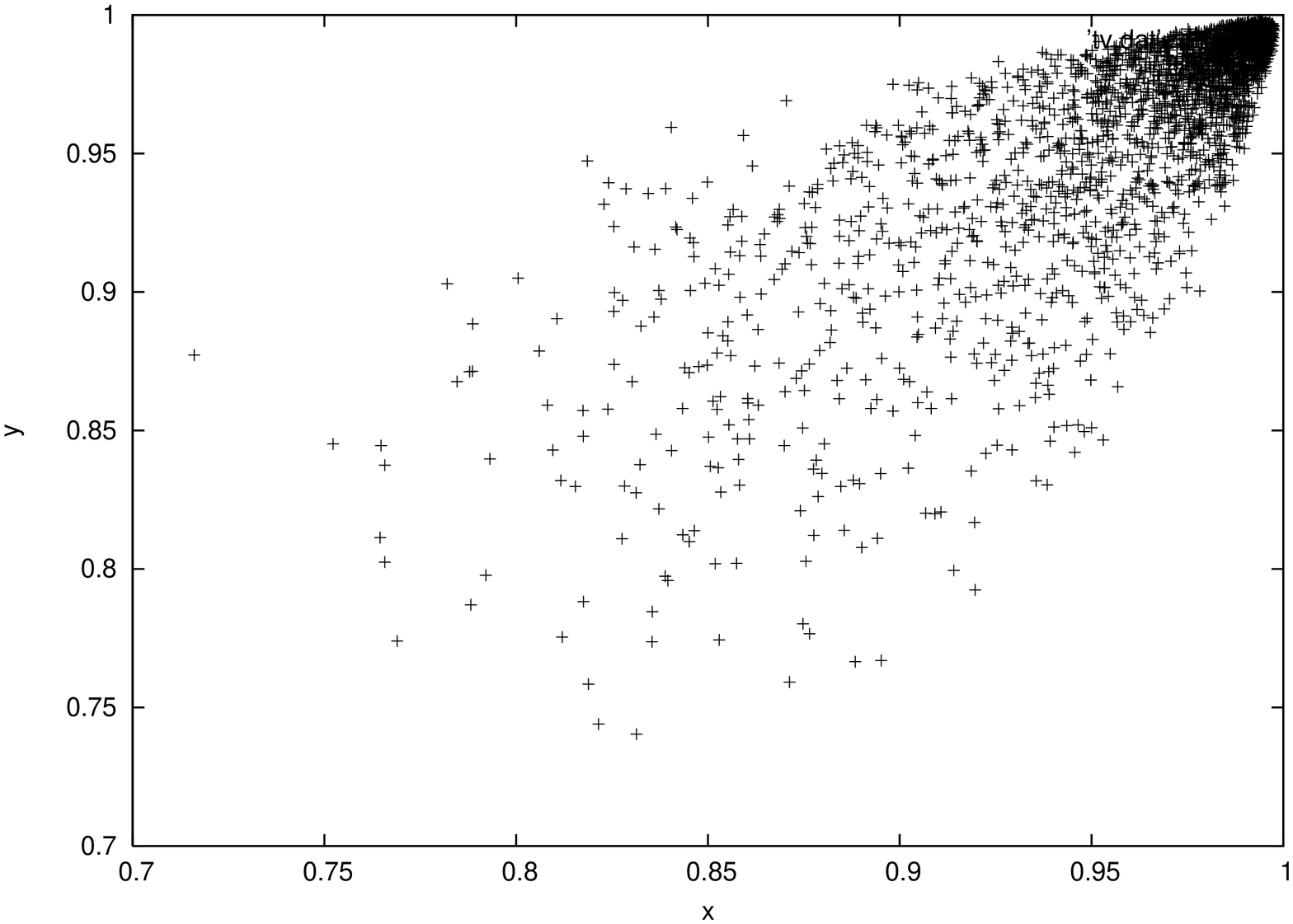,%
width=2.5in,height=3in,angle=270}
\caption{Phase space and distribution of hardest emissions for axial (left) and vector(right) currents with $\rho=0.1225$.}
\label{fig:pa}
\end{figure} 
\begin{figure}[!ht]
\hspace{0.5cm}
\psfig{figure=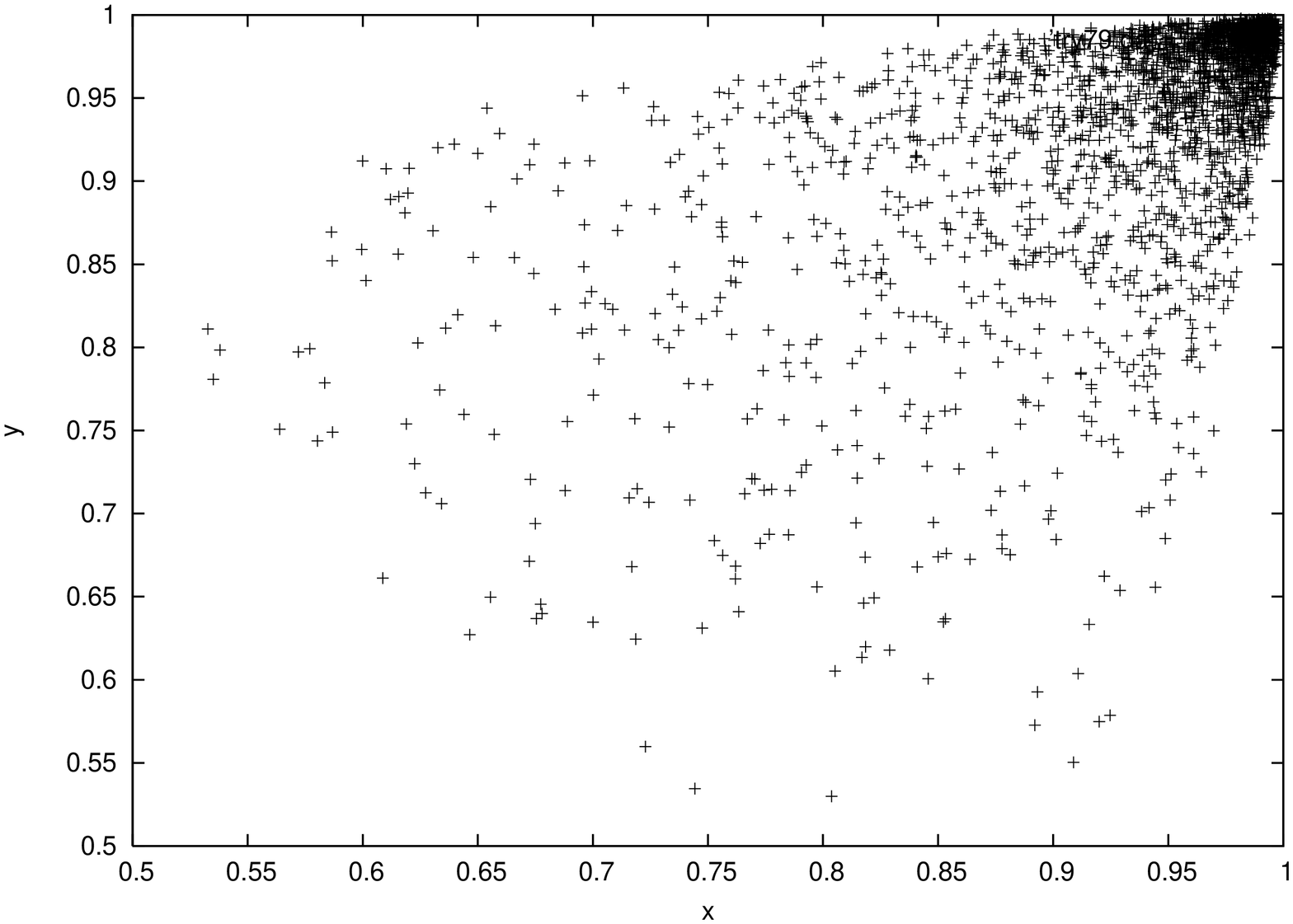,%
width=2.5in,height=3in,angle=270}
\psfig{figure=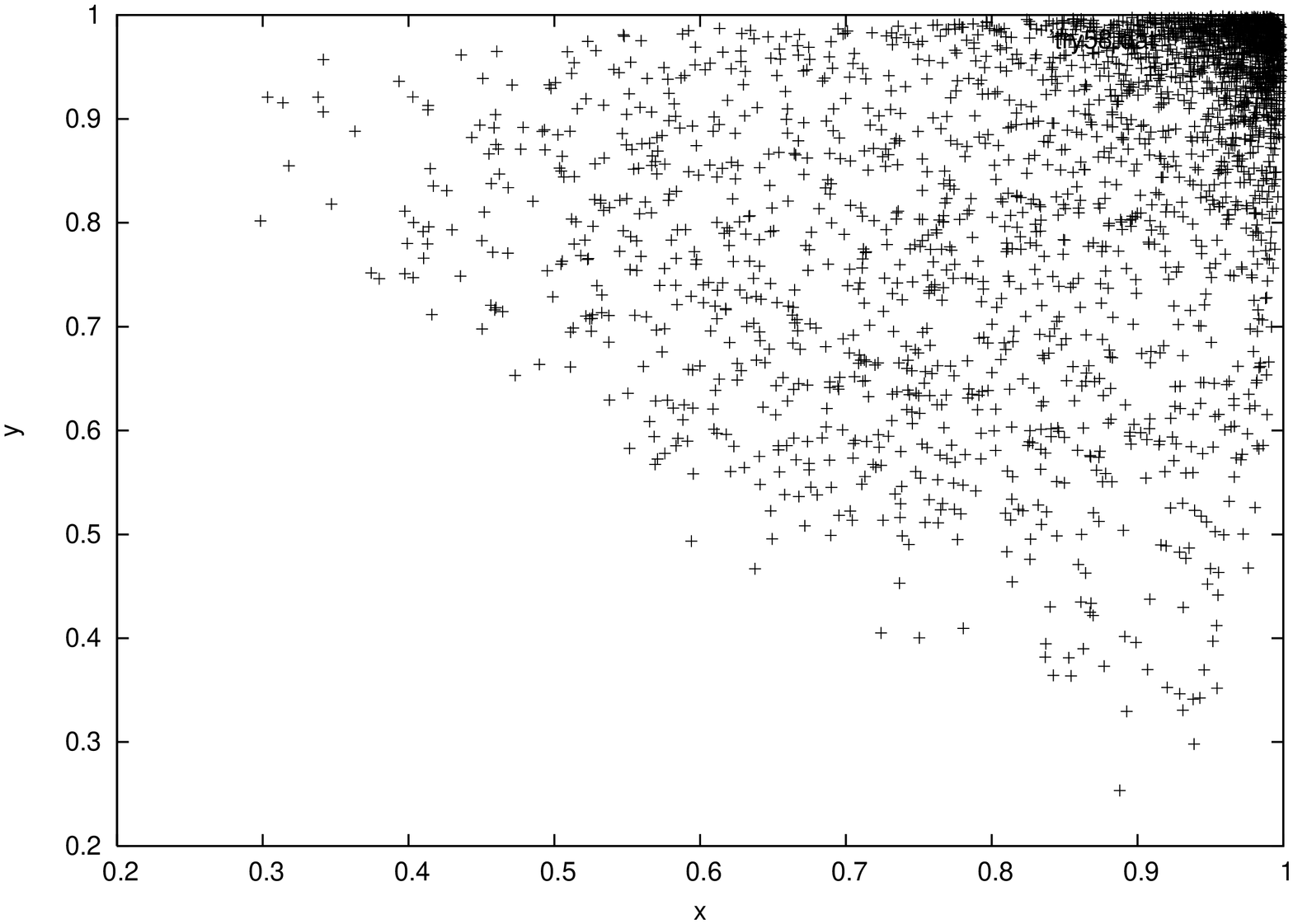,%
width=2.5in,height=3in,angle=270}
\caption{Phase space and distribution of hardest emissions for an axial current with
  $\rho=0.0625$ (left) and $\rho=0.01361$ (right).}
\label{fig:pa}
\end{figure} 
  
%
 The procedure was repeated for $\sqrt{s}=700$ GeV, $\rho=0.0625$ and $\sqrt{s}=1500$ GeV, $\rho=0.01361$ and the corresponding plots are shown below.
As can be seen, the method is stable as $\rho \rightarrow 0$. This is not surprising
because the upper bound function $V(x,\kappa)$ in (\ref{eq:V}) is stable as $\rho \rightarrow 0$ and
tends to $W^{'}$, the true value of the integrand in this limit.
\clearpage
\section{Hardest emission generation: Decays}
\begin{figure}
\begin{center}
\psfig{figure=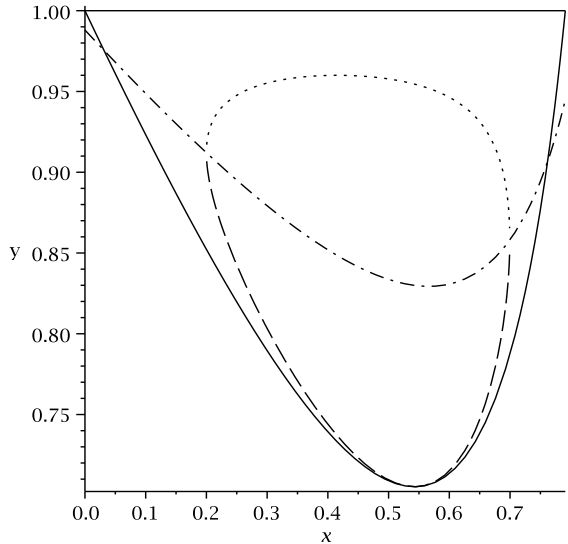,%
width=3.03in,height=2.93in,angle=0}
\end{center}
\caption{Phase space(solid), $y^{'}$(dot-dash) and solutions $y_1$ (dots) and $y_2$ (dashes).}
\label{fig:soln}
\end{figure}

\begin{figure}
\begin{center}
\psfig{figure=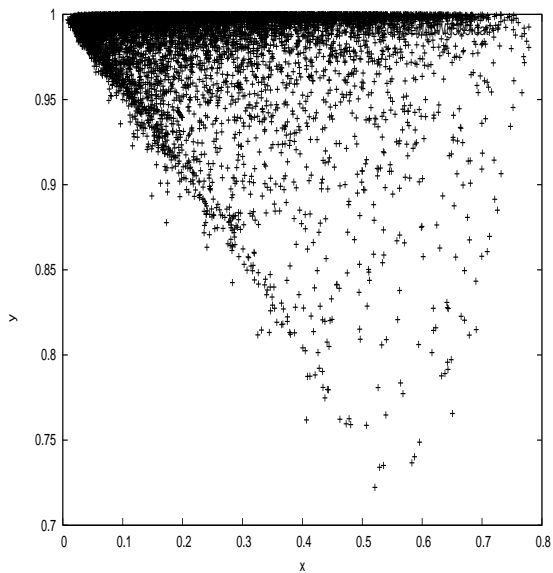,%
width=3.03in,height=2.93in,angle=270}
\end{center}
\caption{Phase space distribution of {\tt POWHEG} events }
\label{fig:psdecay}
\end{figure}

In addition to gluon emission in top production, we also studied the emission in its
decay,
\begin{equation}
\label{decay}
t(p_1) \rightarrow W^{+}(w_1)b(r_1)g(k) \;.
\end{equation} 
The procedure for generating the hardest emission in this case follows the same lines as
discussed in Section \ref{POWHEG}. We parameterize the phase space for the decay in terms of variables $x$ and $y$
defined as
\begin{eqnarray}
y&=&\frac{2w_{1} \cdot p_1}{m_{\rm t}^{2}}-a \nonumber \\
x&=&\frac{2k \cdot p_1}{m_{\rm t}^{2}} 
\end{eqnarray}
where $a=m_{w}^{2}/m_{\rm t}^{2}$ with $m_w$ and $m_{\rm t}$ the masses of the $W$ boson and top
quark respectively. $(y+a)/2$ and $x/2$ are the energy fractions of the $W$ boson and gluon
in the top frame. Therefore the corresponding energy fraction of the $b$ quark in this frame is given by
\begin{eqnarray}
\frac{x_b}{2}=\frac{2-y-a-x}{2} \;.
\end{eqnarray}
In this paper, we neglect the $b$ mass and work in the narrow-width
approximation so that the top quarks and $W$ boson are on-shell.
The $t \rightarrow Wbg$ differential decay rate is given by:
\begin{equation}
\label{decayme}
 \frac{1}{\Gamma_0}\frac{d^2\Gamma}{dxdy}=\frac{\alpha_S}{\pi} \frac{C_F}{(1-y)x^2}\left
  [x-\frac{(1-y)(1-x)+x^2}{1-a}+x\frac{(y+x-1)^2}{2(1-a)^2}+\frac{2a(1-y)x^2}{(1-a)^2(1+2a)}\right] \,,
\end{equation}
where $\Gamma_0$ is the leading order decay rate.
The phase space limits for the decay are:
\begin{eqnarray}
\frac{ax}{1-x}+(1-x) < y < 1 \,,\nonumber\\
0 < x < 1-a \;.
\end{eqnarray}
Working in the rest frame of the top quark where the parton shower is formulated in {\tt
  Herwig++}, we identify the splitting axis corresponding to the original $b-W$ boson axis and
therefore the relative transverse momentum for gluon emission is:
\begin{equation}
k_T(x,y) = m_{\rm t} \sqrt {\frac{(1-y)(y+x(2-y-a)-x^2-1)}{(y+a)^2-4a} } \;.
\end{equation}
Now defining a dimensionless variable $\kappa = \frac{k_{T}^{2}}{m_{\rm t}^2}$, we find that
in analogy to the production case, there are $2$ solutions for $y$ for each value of
$x$ and $\kappa$.
\begin{equation}
y_{1,2}=\frac{x^2+ax+2-3x-2a \kappa \pm \sqrt{(x^2-4\kappa(1+a))(x-1)^2+4a \kappa(4 \kappa+1-a)+x^2(a+2x-2)}}{2(\kappa+1-x)}\;.
\end{equation}
These solutions may be identified with either initial state gluon emission from the top
quark ($y_2$) or final state radiation from the bottom quark ($y_1$). A plot of the phase space and the $2$ solutions for
$\kappa=0.01$ is shown in Figure \ref{fig:soln}.
We then construct the modified Sudakov form factor for the generation of the hardest
emission. The exponent of the form factor is given by
\begin{equation}
\label{Wdec}
\int W(x,y) \Theta (k_T(x,y)-p_{T})dxdy = \int ^{x_{\rm max}}_{x_{\rm min}} dx
\int^{\kappa_{\rm max}}_{\kappa} d \kappa \frac{\alpha_S(\kappa m_{\rm t}^2)C_F}{\pi} \frac
  {dy}{d \kappa} W(x, \kappa) \,,
\end{equation}
where $W(x,\kappa)$ is the differential cross-section (\ref{decayme}) and $\frac{dy}{d
  \kappa}$ is the Jacobian for the change of variables from $y$ to $\kappa$. Note that $\kappa = \kappa_{\rm
  max} = \frac{(1-\sqrt{a})^2}{4}$ when the $W$ boson is at rest and $x=1-\sqrt{a},
y=2\sqrt{a}$. For a given $\kappa$, $x_{\rm min}$ and $x_{\rm max}$ are also given by
\begin{eqnarray}
x_{\rm min}&=&2\sqrt{\kappa} \nonumber \\
x_{\rm max}&=&1-a-2\sqrt{\kappa a} \;.
\end{eqnarray}
To make the integral simpler, we again look for an upper bound
$V^{'}(x,\kappa)$ on the integrand as we did for the production case.  To do this we replace the Jacobian with the simpler expression,
\begin{equation}
\frac{dy^{'}}{d\kappa}=\frac{d}{d \kappa}\left(\frac {x^2-3x-2 \kappa a+2+xa}{2(\kappa+1-x)} \right) =
\frac{-a}{\kappa+1-x} - \frac{x^2-3x-2 \kappa a+2+xa}{2(\kappa+1-x)^2} \,,
\end{equation}  
where $y^{'}$ lies in between $y_{1}$ and $y_2$ and is indicated in Figure
\ref{fig:soln}.
 We also overestimate the differential cross-section by replacing (\ref{decayme}) with
\begin{equation}
U(x,y) = N_{\kappa}\frac{\alpha_S C_F}{\pi} \frac{1-a}{2x^2(1-y^{'})} \,, 
\end{equation}
where $N_{\kappa}$ is a normalisation factor dependent on $\kappa$ and is chosen such that
$V^{'}=U\frac{dy^{'}}{d \kappa}$ is greater than the integrand in (\ref{Wdec}). The
$N_{\kappa}$ values are given in Table \ref{tab:kapdec} for the $2$ solutions.
\begin{table}[!ht]
\centering
\begin{tabular}{llll} \hline
Range of $\kappa$ & $N_{\kappa}(y_{1}) \times 10^{4}$&$N_{\kappa}(y_{2})\times 10^{4}$\\
\hline
\hline                      
$0.01-0.0737$  & $0.005$ & $0.006$ \\ 
$0.005-0.01$   & $0.0175$& $0.02$ \\ 
$0.001-0.005$ & $0.03$ & $0.045$   \\ 
$0.0001-0.001$ & $0.08$ & $0.12$    \\ 
$0.00005-0.0001$ & $0.2$ &$0.2$   \\ 
$0.000025-0.00005$ & $0.3$& $0.2$\\ 
$0.0000075-0.000025$ &$1.0$ & $0.9$   \\ 
$0.000005-0.0000075$ & $2.0$ &$0.9$    \\ 
$0.0000025-0.000005$ & $3.0$ &$0.9$    \\ 
$0.0000013-0.0000025$ & $6.0$ &$0.9$    \\ 
\hline
\hline
\\
\end{tabular}
\caption{$N_{\kappa}$ for different values of $\kappa$}
\label{tab:kapdec}
\end{table}
The lower limit on $\kappa$ is $1.3 \times 10^{-6}$ and was set by choosing $k_T=\Lambda_{\rm
  QCD}=0.2$ GeV thus setting a lower bound on the transverse momentum. We then generate
the values of $\kappa$ and distribute $x$ and $y$ according to the true differential
(\ref{decayme}) using vetoes as described for the production case in Sections \ref{POWHEG}. Figure \ref{fig:psdecay} shows the phase space distribution obtained.
\section{Spin Correlations and the distribution of Born variables}
\label{spin}
In \cite{Frixione:2007zp}, it was observed that the lepton matrix element for the
production process 
\begin{eqnarray} 
\label{lme}
e^+(p)+e^-(q) &\rightarrow& t(p_1)+\bar{t}(p_2)+g(p_3) \rightarrow W^+(w_1)+b(r_1) +
W^-(w_2)+b(r_2)+g(p_3) \nonumber \\
&\rightarrow& l^+(k_1)+\nu (x_1)+b(r_1)+l^-(k_2)+\nu (x_2)+\bar{b}(r_2)+g(p_3)
\end{eqnarray}
 is bounded from
above in the narrow width approximation by the undecayed matrix element obtained by eliminating the decay products
i.e. $W^+, W^-, b,\bar{b}$ and putting the parent particles i.e. $t, \bar{t}$ on-shell,
multiplied by a
process dependent constant. We can then use the undecayed matrix elements to perform
computer-intensive tasks such as event generation and finally, by using the hit-and-miss method,
replace the parent particles with their decay products. This procedure is outlined below:
\begin{enumerate}
\item
Evaluate the undecayed matrix elements which are proportional to the upper bounds on the lepton matrix
elements. Generate hard events using the {\tt POWHEG} method described above with the top
and anti-top quarks in the final state. 
\item
For each event, generate the decay products and their four-momenta according to the phase space.
\item
Evaluate the leptonic decay matrix element for each event. If the decay matrix element divided by the corresponding
upper bound is less than a random number $r$ between $0$ and $1$, throw away
the decay momenta and return to step 2.
\item
Otherwise, replace the top and anti-top momenta with the decay momenta and shower the
event.
\end{enumerate}
\subsection{Undecayed matrix elements}
\label{ume}
At the ILC, the electron and positron beams will be polarized i.e. either $e^-_Le^+_R$ or
$e^-_Re^+_L$ where the subscripts $L$ and $R$ represent the left-handed and right-handed
helicity states respectively. The corresponding undecayed  matrix
element for $e^-_Le^+_R$ annihilation is:
\begin{eqnarray}
\label{me}
&&\tilde {M}(e^-_L(p)e^+_R(q) \rightarrow t_{s_t}(p_1)\bar{t}_{s_{\bar{t}}}(p_2) g(p_3))=\nonumber \\
&&[\bar{v(q)}\gamma_{\mu L}u(p)]\bar{u}(p_1,s_t)\left[\frac{1}{2p_2 \cdot
    p_3}\left(\frac{a_{LL}}{2}\gamma^{\mu}_{L}+\frac{a_{LR}}{2}\gamma^{\mu}_{R}
  \right)(-\hat{p_2}-\hat{p_3}+m_{\rm t}) \gamma_{\nu}\right.\nonumber \\
&&+\left.\frac{1}{2p_1 \cdot
    p_3}\gamma_{\nu}(\hat{p_1}+\hat{p_3}+m_{\rm t}) \left(\frac{a_{LL}}{2}\gamma^{\mu}_{L}+\frac{a_{LR}}{2}\gamma^{\mu}_{R}
  \right)\right]T^av(p_2,s_{\bar{t}})\epsilon_a^{\nu}(p_3)
\end{eqnarray}
where $\epsilon$ is the polarization vector of the gluon, $T^a$ is the colour matrix,
$\hat {p}=p^{\mu}\gamma_{\mu}$ and $\gamma^{\mu}_{R/L}=\gamma^{\mu}(1 \pm \gamma_5)/2$. For $e^-_Re^+_L$ annihilation, interchange $L,R$ in the
above equation.  $s_t,s_{\bar{t}}$ are the spin vectors of the top and
anti-top quarks respectively and satisfy the relations:
\begin{eqnarray}
s_t \cdot p_1 &=& 0 \nonumber \\
s_{\bar {t}} \cdot p_2 &=& 0 \nonumber \\
s_t \cdot s_t &=& -1 \nonumber \\
s_{\bar {t}} \cdot s_{\bar {t}} &=& -1
\end{eqnarray}
The massive spinors $u(p,s),v(p,s)$ are given in terms of
the massless spinors $u(p),v(p)$ by
\begin{eqnarray}
u(p,\uparrow) &=& \frac{1+\gamma_5 \hat{s}}{2}u(p)\nonumber \\
u(p,\downarrow) &=& \frac{1-\gamma_5 \hat{s}}{2}u(p)\nonumber \\
v(p,\uparrow) &=& \frac{1+\gamma_5 \hat{s}}{2}v(p)\nonumber \\
v(p,\downarrow) &=& \frac{1-\gamma_5 \hat{s}}{2}v(p)\nonumber \\
\end{eqnarray}
The coupling constants $a_{IJ}$ are given by
\begin{eqnarray}
a_{IJ}=\frac{e^2g}{s}\left[-Q_t+Q_e^IQ_t^J\frac{1}{\sin^2\theta_W}\frac{s}{s-M_Z^2+iM_Z\Gamma_Z}\right]
\end{eqnarray}
where $M_Z$ is the $Z$ boson mass, $\Gamma_Z$ is the width of the $Z$ boson, $\theta_W$ is
the Weinberg angle, $Q_t$ is the electric charge of the top in units of the electric
charge $e$, $g = \sqrt{4 \pi \alpha_S}$ and $s$ is the center of mass
energy squared. The couplings to the $Z$ boson are given by
\begin{eqnarray}
Q_e^L&=&\frac{2 \sin^2 \theta_W-1}{2 \cos \theta_W} \nonumber \\
Q_e^R&=&\frac{\sin^2 \theta_W}{\cos \theta_W} \nonumber \\
Q_t^L&=&\frac{3-4 \sin^2 \theta_W}{6 \cos \theta_W} \nonumber \\
Q_t^R&=&-\frac{2 \sin^2 \theta_W}{3 \cos \theta_W} 
\end{eqnarray}

In Section \ref{sec:Vxy}, we distributed our events
according to the vector and axial vector current matrix elements separately
using the {\tt POWHEG} method. To obtain a full unpolarized distribution we can select
events from either current distribution according to their contributions to the full
cross-section given below.
\begin{eqnarray}
\sigma&=&3\beta(1+2\rho)\left(1+c_1\frac{\alpha_S}{\pi}\right)\sigma_{VV}+3\beta^3\left(1+d_1\frac{\alpha_S}{\pi}\right)\sigma_{AA} \nonumber \\
\sigma_{VV}&=&\frac{4 \pi
  \alpha_{\rm em}^2}{s}\left[Q_t^2-2Q_tV_eV_t\chi_1(s)+(A_e^2+V_e^2)V_t^2\chi_2(s) \right]\nonumber \\
\sigma_{AA}&=&\frac{4 \pi \alpha_{em}^2}{s}\left[(A_e^2+V_e^2)A_t^2\chi_2(s) \right] \,,
\end{eqnarray}
where $\beta=\sqrt{1-\rho}$, $\alpha_{\rm em}$ is the electromagnetic coupling, $A_t, A_e$ and $V_t, V_e$ are the axial and vector coupling constants of
the top $t$ and electron $e$ to the $Z$ boson and $c_1=3.5$ and $d_1=2.25$ are the QCD correction coefficients
defined at $m_t=175$ GeV and $\sqrt{s}=500$ GeV i.e. $\rho=0.1225$ \cite{Jersak:1981sp}. $\chi_1(s)$
and $\chi_2(s)$ are given by
\begin{eqnarray}
\chi_1(s)&=&\kappa \frac{s(s-M_Z^2)}{(s-M_Z^2)^2+\Gamma_Z^2M_Z^2} \nonumber \\
\chi_2(s)&=&\kappa^2 \frac{s^2}{(s-M_Z^2)^2+\Gamma_Z^2M_Z^2} \nonumber \\
\kappa &=& \frac{\sqrt{2}G_FM_Z^2}{16 \pi \alpha_{\rm em}} \,,
\end{eqnarray}
where $G_F$ is the Fermi constant and $M_Z$ and $\Gamma_Z$ are the mass and decay width of
the $Z$ boson respectively.

Explicit expressions for the Born, virtual and real polarization
dependent squared matrix elements for the production process are given in \cite{Schmidt:1995mr} in terms of the energy fractions
$x, y$ of the top and anti-top quarks and the polar angle and azimuthal angles orienting the $t\bar{t}g$ plane relative to the $e^+e^-$ beam axis. 
For each initial polarization, we then assign final-state polarizations to each event in
proportion to the squared matrix elements and distribute the polar and azimuthal angles of the
top/anti-top pairs accordingly using well-known Monte Carlo techniques.
\subsection{Decay matrix elements}
\label{leptonme}
Next, we investigate the decays of the top and anti-top pair. The leptonic matrix elements
for the process in (\ref{lme}) are dependent on the spins of the top and anti-top
quark. This dependence can be written in the form of a decay density matrix. The
decay density matrix $\rho_{\lambda, \lambda^{'}}$, for an on-shell top quark is given by 
\begin{eqnarray}
&&\rho_{ \lambda, \lambda^{'}} = \frac{4g_w^4 V_{tb}^2}{(w_1^2-m_w^2)^2+(m_w \Gamma_W)^2}
\times \nonumber \\
&&\left[\begin {array} {clcr}
(r_1 \cdot x_1)(p_1 \cdot k_1)-(s_t \cdot k_1)(r_1 \cdot x_1)m_{\rm t} &
-(k_1 \cdot n)(x_1 \cdot r_1)m_{\rm t}-i\epsilon(p_1,k_1,s_t,n)(x_1 \cdot r_1)\nonumber\\ 
-(k_1 \cdot n)(x_1 \cdot r_1)m_{\rm t}+i\epsilon(p_1,k_1,s_t,n)(x_1 \cdot r_1) &
(r_1 \cdot x_1)(p_1 \cdot k_1)+(s_t \cdot k_1)(r_1 \cdot x_1)m_{\rm t} 
\end{array} \right] 
\end{eqnarray}
where $\lambda, \lambda^{'}$ are spin labels, $s_t$ is the top spin vector, $n$ is a spacelike vector perpendicular to $s_t$ and
$p_1$ and $m_w,\Gamma_W$ are the mass and width of the $W$ boson respectively. In this
paper we work in the helicity basis for which the top quark spin is defined along its
direction of motion. A similar matrix can be derived for $\bar{t}$ decay. The spin-specific decay
matrix elements are therefore of the form:
\begin{equation}
S_{\lambda_t \lambda_{\bar{t}} \lambda_t^{'} \lambda_{\bar{t}}^{'}}= \tilde{M}_{\lambda_t \lambda_{\bar{t}}} \rho_{ \lambda_t \lambda_{\bar{t}}^{'}} \rho_{ \lambda_{\bar{t}}
  \lambda_{\bar{t}}^{'}} \tilde{M}^*_{\lambda_t^{'} \lambda_{\bar{t}}^{'}}
\end{equation}
where $\lambda_t, \lambda_{\bar{t}}$ are spin labels for the top and anti-top respectively
and $\tilde {M}$ is the matrix element for the undecayed process introduced in Section \ref{ume}.
 By diagonalizing the density matrix,
we can obtain the largest possible value of the matrix elements and hence the upper
bound. An explicit computation gives this upper bound $\mid M^{t}_{ub}\mid ^2$ on the top decay as \cite{Frixione:2007zp},
\begin{equation}
\label{ub}
\mid M^{t}_{ub}\mid ^2=\frac{4g_w^4 \mid V_{tb} \mid^2 (r_1 \cdot x_1)(p_1 \cdot k_1)}{[(w_1^2-m_w^2)^2+(m_w
  \Gamma_W)^2][(p_1^2-m_{\rm t}^2)^2-(m_{\rm t} \Gamma_t)^2]}\mid \tilde {M} \mid ^2
\end{equation}
where $ \mid\tilde {M} \mid ^2$ is the  undecayed matrix element for unpolarized
$t\bar{t}g$ production. A similar expression $\mid M^{\bar {t}}_{ub}\mid ^2$ can be
obtained for the decay of the top anti-quark by interchanging the labels 1 and 2 and $t$ and $\bar{t}$ in (\ref{ub}). Hence the full upper bound can be written as:
\begin{equation}
\mid M^{t\bar{t}}_{ub}\mid^2=\frac{\mid M^{t}_{ub}\mid ^2\mid M^{\bar {t}}_{ub}\mid ^2}{\mid \tilde {M} \mid ^2}
\end{equation}
 
Having obtained the decay matrix elements and their upper bounds, we then proceed to
generate events with leptons in the final state as outlined at the beginning of this
section.

For the {\tt POWHEG} decays, we apply the same method where in this case the undecayed
matrix elements $|\tilde{M}|$ are the leading order matrix elements for the process
\begin{equation}
e^{+}+e^{-} \rightarrow t + \bar{t} \;.
\end{equation}
We then use the next-to-leading order decay matrix for which the helicity amplitudes can
be found in \cite{Schmidt:1995mr}. These are given in terms of the polar and azimuthal
angles of the decay w.r.t the top/anti-top axis and we distribute them as described for the production process in Section \ref{ume}. Note that in this
case, we generate two decay gluons, one each from the top and anti-top quark.

In addition, we also consider {\tt POWHEG} radiations in both the production and decay process
by independently generating the emission and distributing the Born variables of the
production process first and then generating the emission and distributing the Born
variables of the decay process to yield three gluons in the final state.

\section{Decay NLO lepton spectra comparisons}
\begin{figure}[!ht]
\vspace{2cm}
\hspace{0.5cm}
\psfig{figure=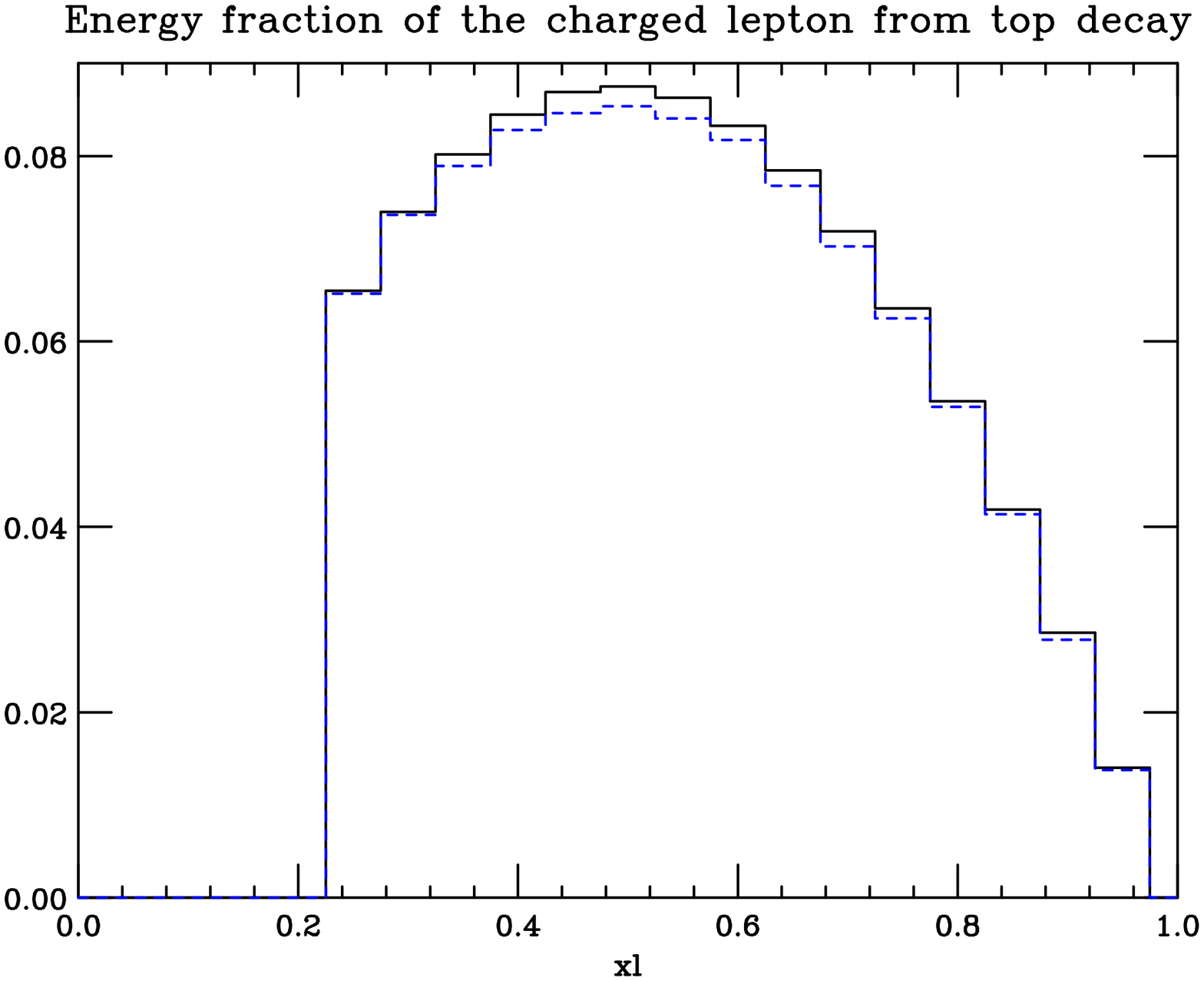,%
width=2.5in,height=3in,angle=90}
\psfig{figure=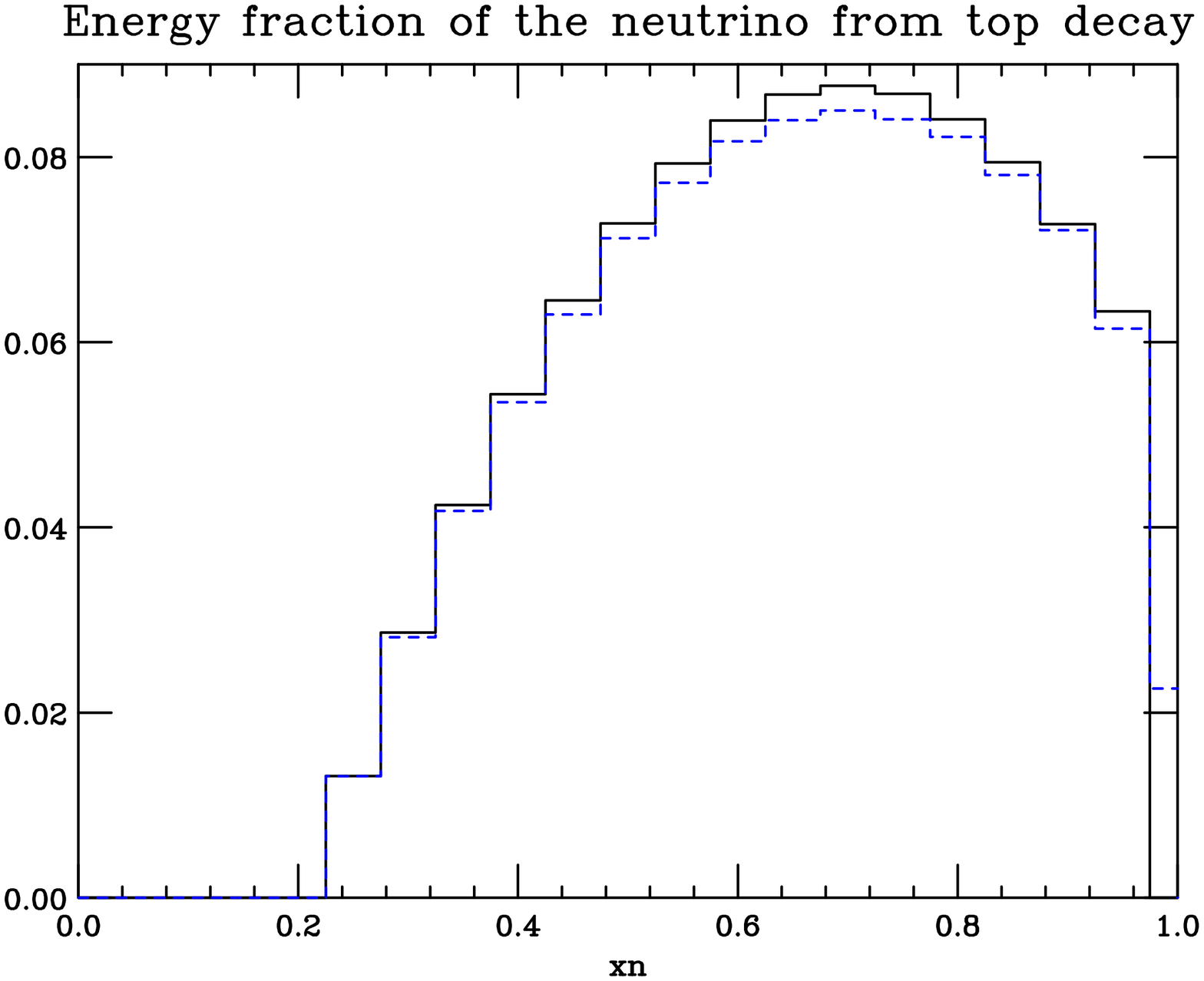,%
width=2.5in,height=3in,angle=90}
\caption{Scaled energy fractions of the charged lepton (left) and neutrino (right) from
  top decay. Black(solid)= Theory, Blue(dashes)= Decay.
}
\label{fig:xlxn}
\end{figure}

Extensive studies have been carried out on the lepton angular and energy distributions from the semi-leptonic
decays of polarized top and anti-top quarks at next-to-leading
order in $\alpha_S$ \cite{Jezabek:1988iv,Czarnecki:1990pe}. 
\begin{eqnarray}
t &\rightarrow& W^{+}+b+g \rightarrow e^{+} + \nu_e+b+g \nonumber \\
\bar{t}&\rightarrow& W^{-}+\bar{b}+g \rightarrow e^{-}+\bar{\nu_e}+\bar{b}+g \;.
\end{eqnarray}
In the top rest frame, we define $\theta$ as the angle between the spin 3-vectors ${\bf
  s_{t},s_{\bar{t}}}$ of the decaying quark and the lepton. We also defined the scaled energies
$x_{l,n}$ of the charged lepton and the neutrino respectively as
\begin{eqnarray}
x_l&=&\frac{2E_l}{m_{\rm t}} \nonumber \\
x_n&=&\frac{2E_n}{m_{\rm t}} 
\end{eqnarray}
where $E_l$ and $E_n$ are the energies of the charged lepton and neutrino in the top rest
frame. In these variables, the NLO double differential distribution of the charged
lepton and neutrino in the decay of a heavy top or anti-top quark with polarization $S$ has been shown to be of the
form
\begin{eqnarray}
\frac{d \Gamma^{l,n}}{dx_{l,n}d \cos \theta}&=&\frac{G_F^{2}m_{\rm t}^5}{32
  \pi^{3}}\left[F_{0}^{l,n}(x_{l,n},a)+S\cos \theta J_{0}^{l,n}(x_{l,n},a) \nonumber
\right. \\ 
&-& \left . \frac{2\alpha_S}{3
    \pi}(F_{1}^{l,n}(x_{l,n},a)+S\cos \theta J_{1}^{l,n}(x_{l,n},a))\right] \,,
\end{eqnarray}
in the narrow width limit for the decay of the $W$ boson. Expressions for $F^{l,n}_{0,1}$ and $J^{l,n}_{0,1}$ can
be found in \cite{Czarnecki:1990pe}.
Integrating over $\cos{\theta}$ gives us the differential energy distribution,
\begin{equation}
\label{xl}
\frac{d \Gamma^{l,n}}{dx_{l,n}}=\frac{G_F^{2}m_{\rm t}^5}{16
  \pi^{3}}\left[F_{0}^{l,n}(x_{l,n},y)-\frac{2\alpha_S}{3
    \pi}F_{1}^{l,n}(x_{l,n},y)\right] \;.
\end{equation}
We compared this theoretical prediction with the distribution obtained from the {\tt
  POWHEG} method before interfacing with the \textsf{Herwig++} parton shower. 
The best fit
distributions shown in Figure \ref{fig:xlxn} were obtained by setting
$\alpha_S$ to $0.1$ in (\ref{xl}).
\section{Truncated Shower}
\label{sec:trunc}
The {\tt POWHEG} method requires the addition of  a `truncated shower' before the
hardest gluon emission in order to simulate the soft radiation distribution \cite{Nason:2006hf}. Due to angular ordering,
the `truncated' radiation  is emitted at a wider angle than the angle of the hardest
emission but at a lower $p_T$. This means the `truncated' radiation does not appreciably degrade
the energy entering the hardest emission and justifies our decision to generate the
hardest emission first.

In \cite{LatundeDada:2006gx}, there is a description of a method to generate a truncated
shower of at most one gluon for the
case of light quark production from $e^+e^-$ annihilation. In this section, we
extend the discussion to top pair production. Below is an outline of how the `truncated
shower' was generated. We will consider the case
in which at most one extra gluon is emitted by the top or anti-top before the hardest
emission. The outline closely
follows the \textsf{Herwig++} parton shower evolution method described in
\cite{Gieseke:2003hm, Gieseke:2003rz} where the
evolution variables $z$, the momentum fractions, and $\tilde{q}$, the evolution scale,
determine the kinematics of the shower. 
\begin{enumerate}[i)]
\item
Having generated the $p_T$ of the hardest emission as discussed in Section \ref{POWHEG} and the energy fractions $x$ and $y$,
calculate the light-cone momentum fractions $z$ and $1-z$ of the partons involved in the hardest
emission. We will assume henceforth that
$x > y$ and that $y$ is the energy fraction of the quark, i.e. the quark is involved in
the hardest emission. Then
\begin{equation}
\centering
\label{eq:z}
z=\frac{ \alpha_b}{ \alpha_b+ \alpha_g} 
\end{equation}
where if we define
\begin{eqnarray}
b&=&\frac{m_{\rm t}^2}{s} \nonumber \\
\lambda&=&\sqrt(1-4b)
\end{eqnarray}
we have
\begin{eqnarray}
\alpha_b&=&\frac{x(1+\lambda)+\sqrt{x^2(1+\lambda)^2-8(b+\kappa)(1+\lambda-2b)}}{2(1+\lambda-2b)}\nonumber \\
\alpha_c&=&\frac{y(1+\lambda)-\sqrt{y^2(1+\lambda)^2-8b(1+\lambda-2b)}}{2(1+\lambda-2b)}\nonumber
  \\
\alpha_g&=& \frac{2}{1+\lambda}-\alpha_b-\alpha_c 
\end{eqnarray}
\item
Next generate the light-cone momentum fraction $z_t$ of the `truncated' radiation within the range
\begin{equation}
\centering
\label{eq:mu}
\frac{m_{\rm t}}{\tilde{q}_i}<z_t<1-\frac{Q_g}{\tilde{q}_i}
\end{equation}
and distributed according to the 
massive splitting function,
$P_{QQ}=C_F\left[\frac{1+z_t^2}{1-z_t}-\frac{2m_{\rm t}^2}{z_t(1-z_t)\tilde{q}^2}\right]$. $\tilde{q}_i$ is the initial evolution scale, i.e. $\sqrt{s}=500$ GeV, and $Q_g$ is a cutoff
introduced to regularize soft gluon singularities in the splitting functions. In this
report, a $Q_g$ value of $0.75$ GeV was used. $z_t$ is the
momentum fraction of the quark after emitting the `truncated' gluon with momentum fraction $1-z_t$.\\
\item
Determine the scale $\tilde{q}_h$ of the hardest
emission from
\begin{equation}
\centering
\label{eq:q}
\tilde{q}_h=\sqrt{\frac{{p_T}^2}{z^2{(1-z)}^2}+\frac{m_{\rm t}^2}{z^2}+\frac{{Q_g}^2}{z{(1-z)}^2}}
\end{equation} \\
\item
Starting from an initial scale $\tilde{q}_i$, the probability of there being an emission
next at the scale $\tilde{q}$ is given by
\begin{equation}
\centering
\label{eq:S}
S(\tilde{q}_i,\tilde{q})=\frac{\Delta(\tilde{q}_c,\tilde{q}_i)}{\Delta(\tilde{q}_c,\tilde{q})}
\end{equation}
where 
\begin{equation}
\centering
\label{eq:dt}
{\Delta(\tilde{q}_c,\tilde{q})}=\exp\left[-\int_{\tilde{q}_c}^{\tilde{q}} 
  \frac{{d\tilde{q}}^{2}}{{\tilde{q}}^{2}}\int dz \frac{\alpha_s}{2\pi}P_{QQ}\Theta(0< p_{T}^{t}< p_{T})\right].
\end{equation}
$\tilde{q}_c$ is the lower cutoff of the parton shower which was set to the default value
of $0.631$ GeV in this
report, $\alpha_s$ is the running coupling constant evaluated at $z(1-z)\tilde{q}$,
$P_{QQ}$ is the $Q \rightarrow Qg$ splitting function and
$p_T$ is the transverse momentum of the hardest emission. The Heaviside function ensures
that the transverse momentum, ${p_{T}^{t}}$ of the truncated emission is real and
is less than $p_T$.
To evaluate the integral in (\ref{eq:dt}), we overestimate the integrands  and apply vetoes
with weights as described in
\cite{Gieseke:2003hm}. With $r$ a random number between $0$ and $1$, we then solve the equation 
\begin{equation}
\centering
\label{eq:r}
S(\tilde{q}_i,\tilde{q})=r
\end{equation}
for $\tilde{q}$. If $\tilde{q} > \tilde{q}_{h}$, the event has a `truncated'
emission. If $\tilde{q} < \tilde{q}_{h}$ , there is no `truncated' emission and the event is showered from the scale of the
hardest emission.\\
\item
If there is a `truncated' emission, the next step is to determine the transverse momentum
$p_{T}^{t}$ of the emission. This is given by 
\begin{equation}
\centering
\label{eq:pT}
p_{T}^{t}=\sqrt{(1-z_{t})^2(z_{t}^2{\tilde{q}}^{2}-m_{\rm t}^2)-z_{t}{Q_{g}}^{2}}\;.
\end{equation}
If ${p_{T}^{t}}^2 < 0$ or  $p_{T}^{t} > p_{T}$ go to ii).\\
\item
We now have values for $z_t$, the momentum fraction of
the quark after the first emission, $p_{T}^{t}$, the transverse momentum of the
first emission, $z$, the momentum fraction of the hardest emission and $p_T$, the transverse momentum of the
hardest emission. We can then reconstruct the momenta of the partons as described in
\cite{Gieseke:2003hm}. The orientation of the quark, antiquark and hardest emission with
respect to the beam axis is determined as explained there for the hard matrix element
correction.\\
\end{enumerate}
In this paper, we consider only truncated emissions in the production process, not in the decay.
\section{Parton shower distributions}
\label{results}
\begin{figure}[!ht]
\hspace{0.5cm}
\psfig{figure=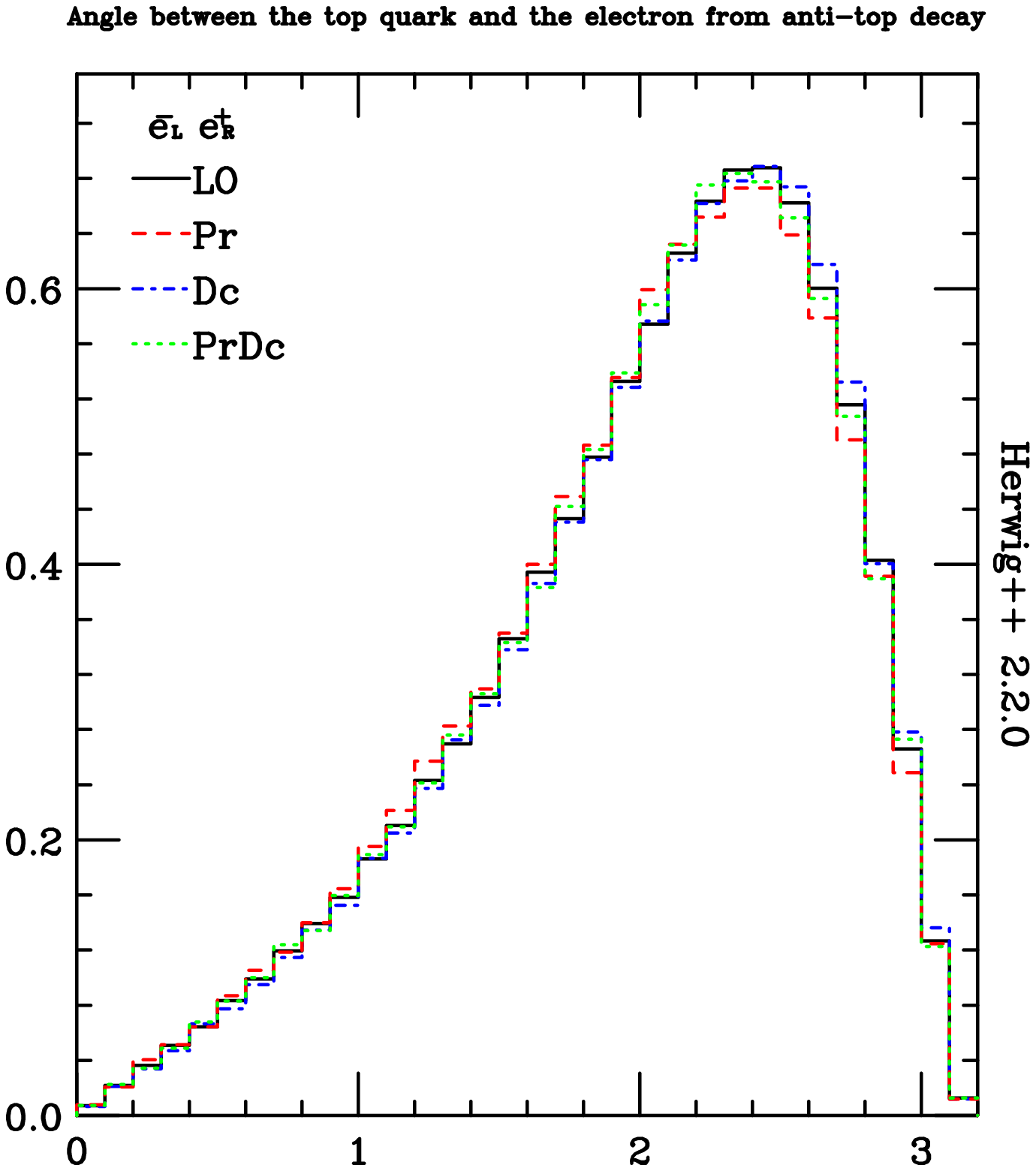,%
width=2.2in,height=3in,angle=90}
\psfig{figure=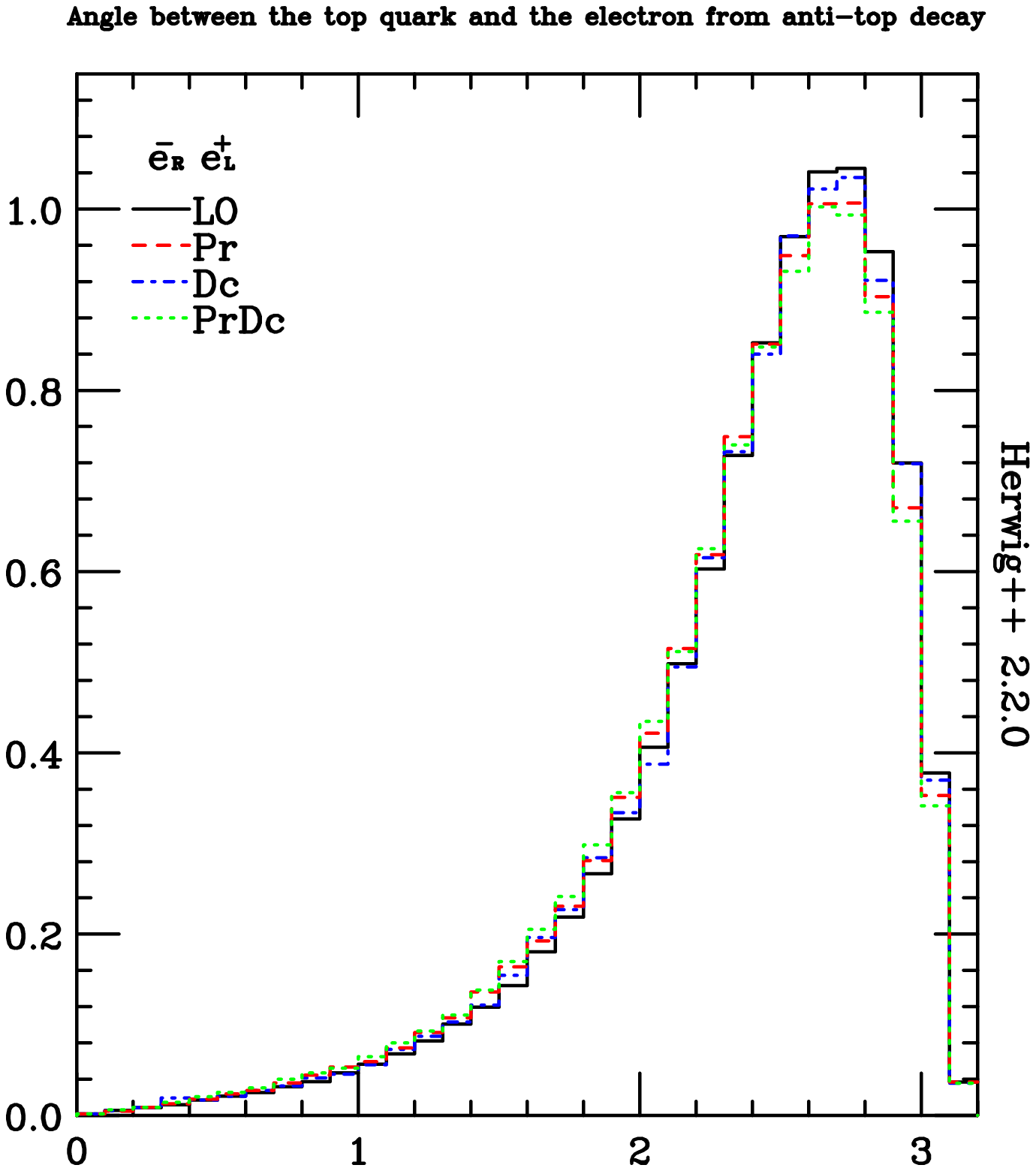,%
width=2.2in,height=3in,angle=90}
\caption{Angle between the lepton from the decay of the top anti-quark and the top quark.}
\label{fig:et}
\end{figure}

\begin{figure}[!ht]
\hspace{0.5cm}
\psfig{figure=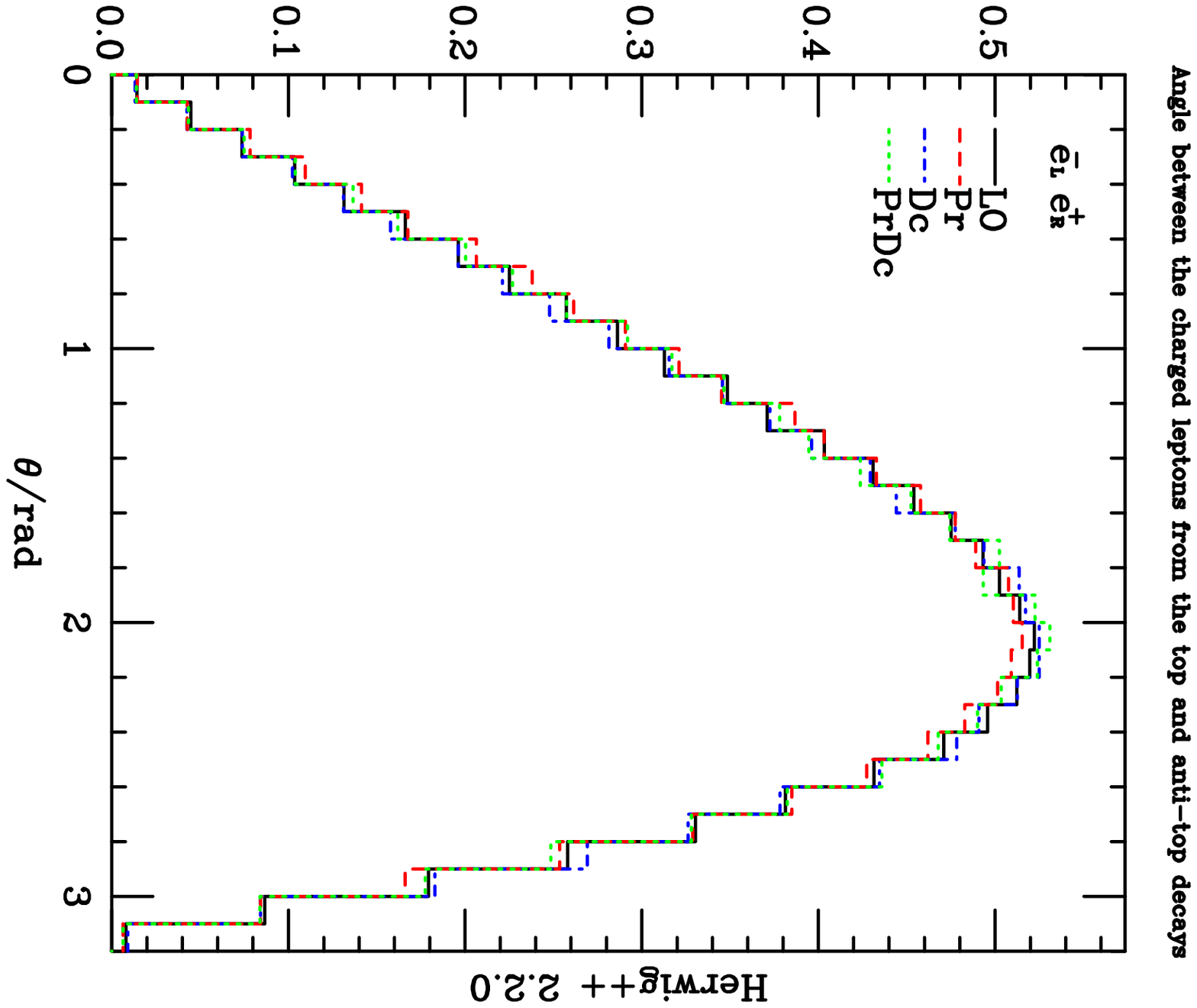,%
width=2.2in,height=3in,angle=90}
\psfig{figure=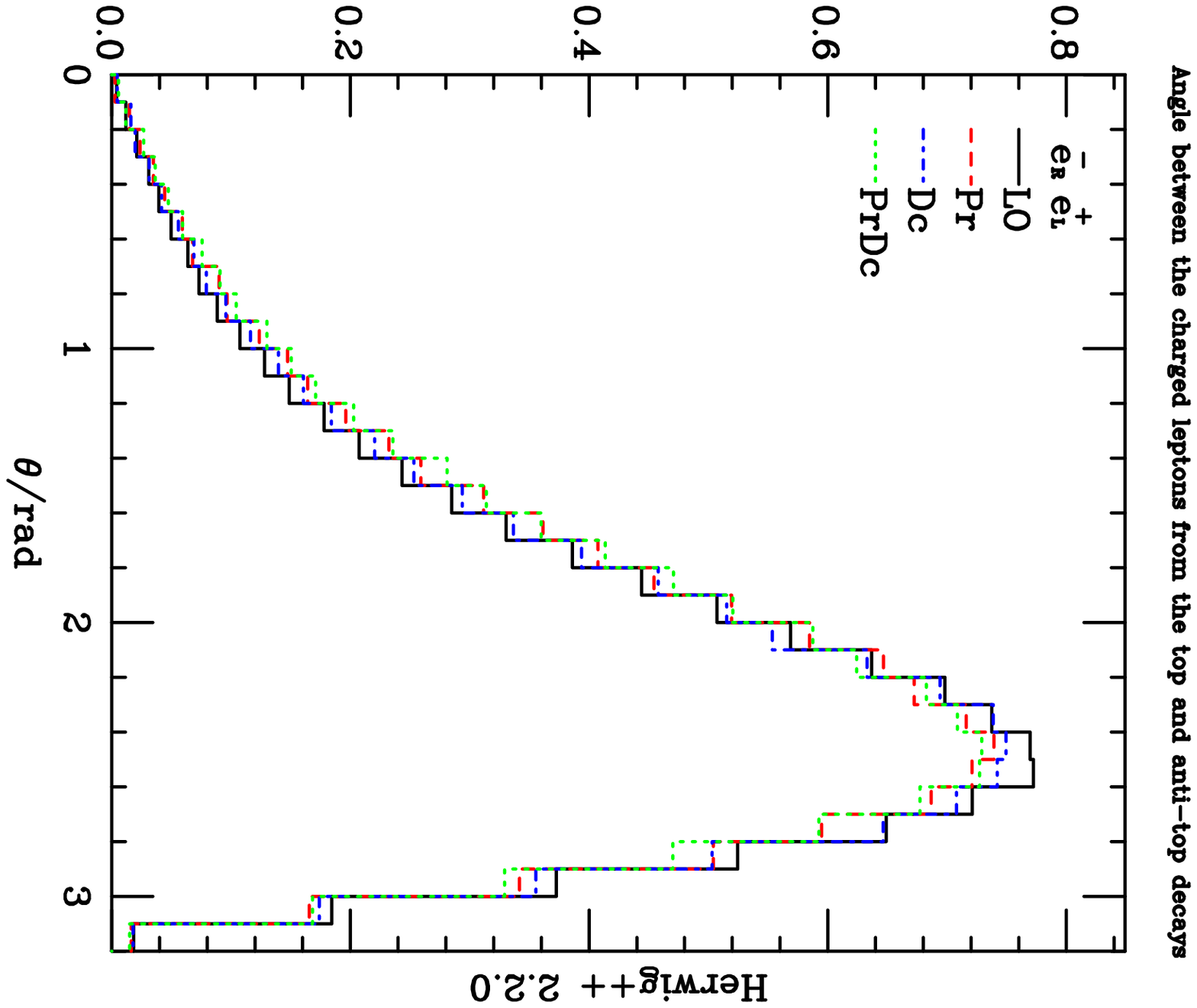,%
width=2.2in,height=3in,angle=90}
\caption{Angle between the lepton and anti-lepton from the decays of the top pairs.}
\label{fig:ee}
\end{figure}

\begin{figure}[!ht]
\vspace{2cm}
\hspace{0.5cm}
\psfig{figure=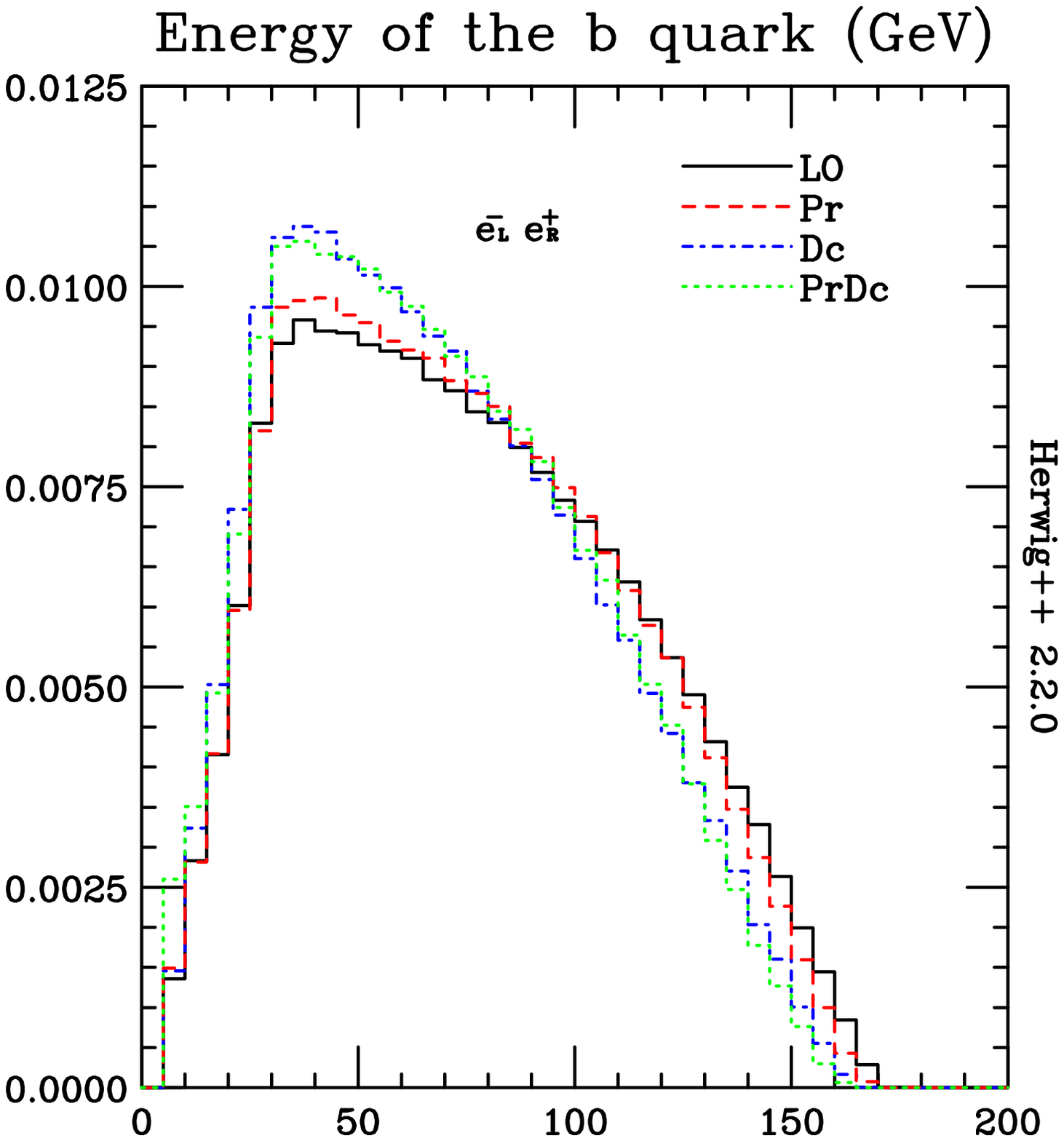,%
width=2.5in,height=3in,angle=90}
\psfig{figure=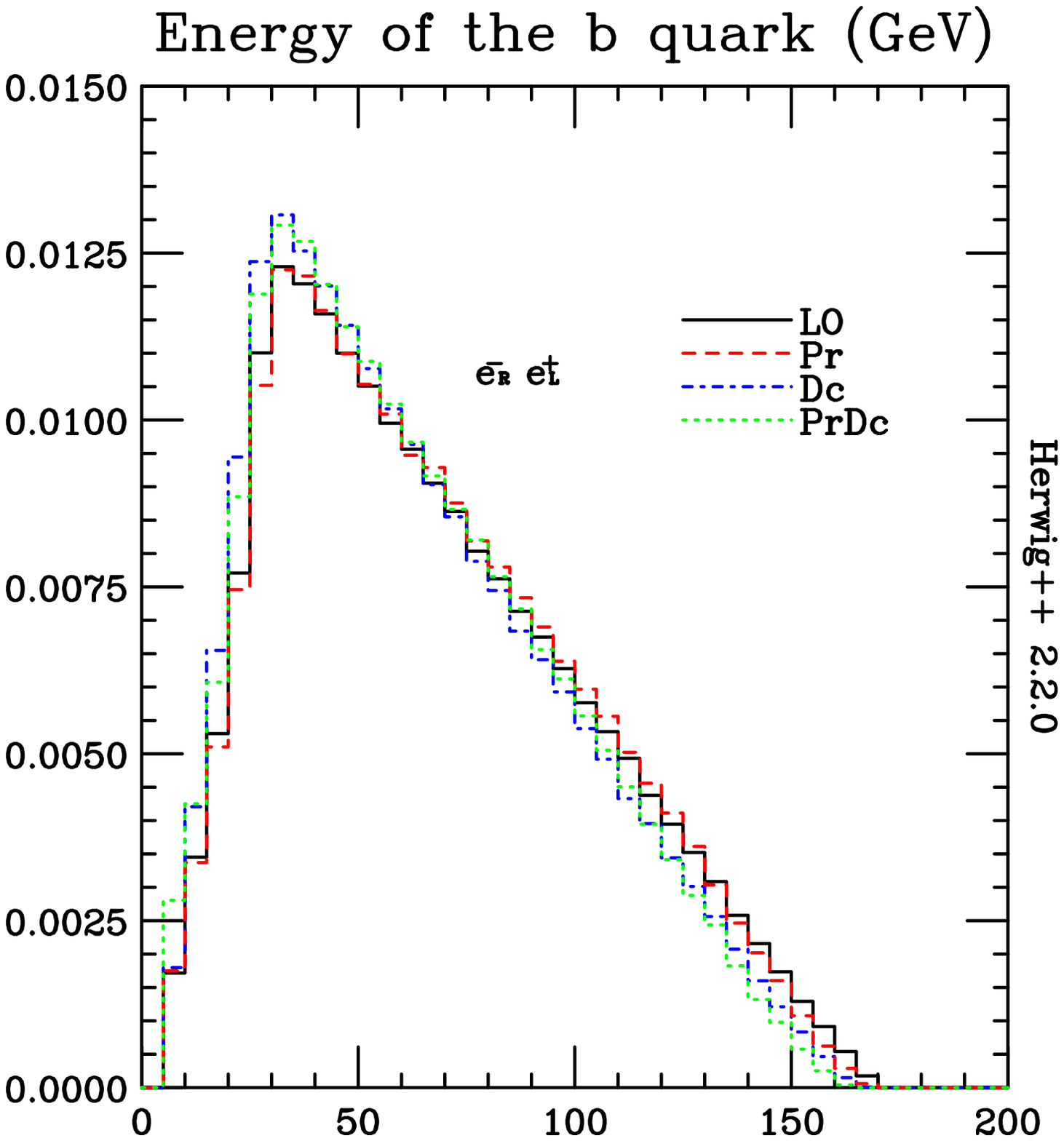,%
width=2.5in,height=3in,angle=90}
\caption{Energy of the b-quark before hadronization.}
\label{fig:benergy}
\end{figure}

\begin{figure}[!ht]
\vspace{2cm}
\hspace{0.5cm}
\psfig{figure=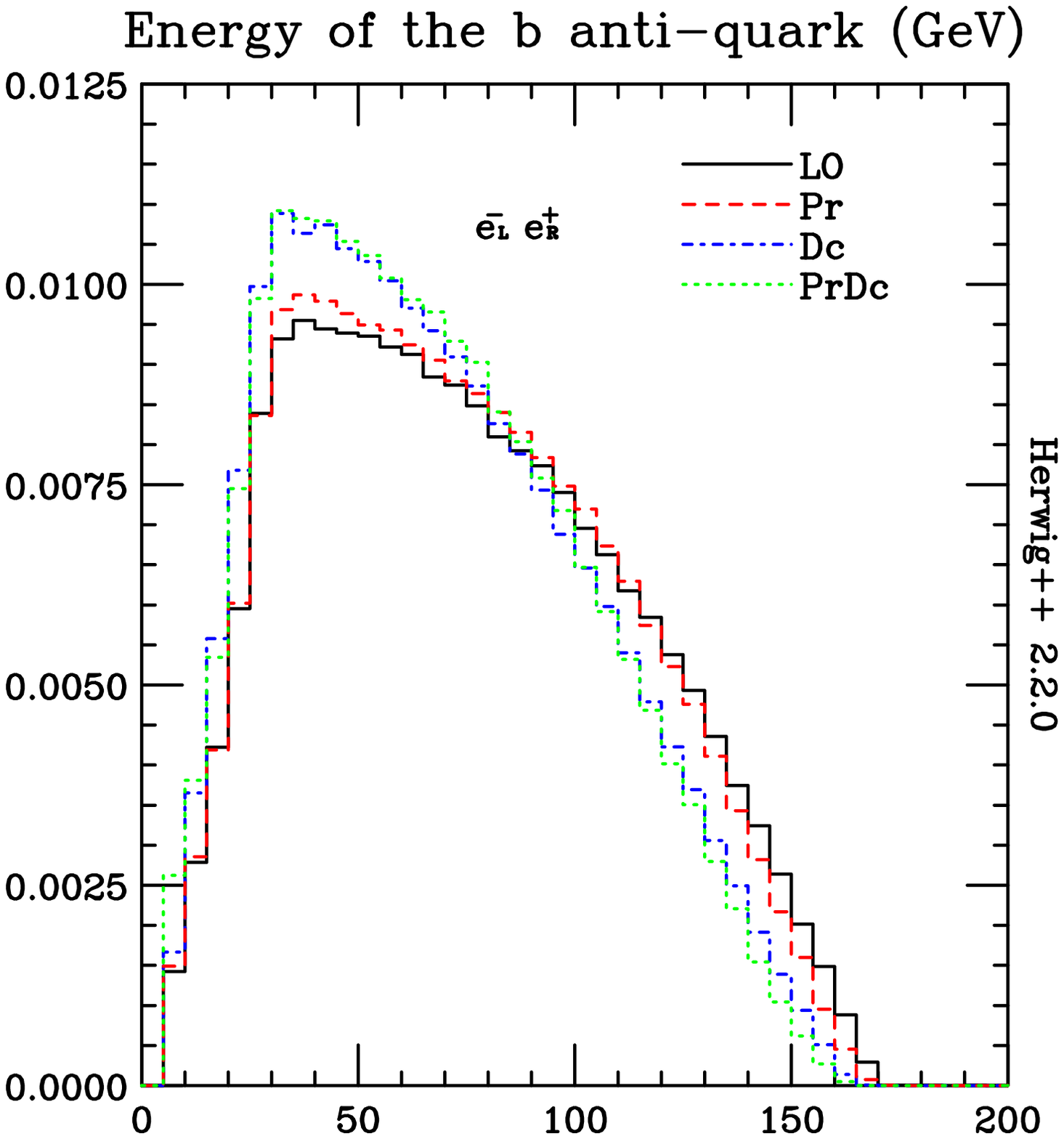,%
width=2.5in,height=3in,angle=90}
\psfig{figure=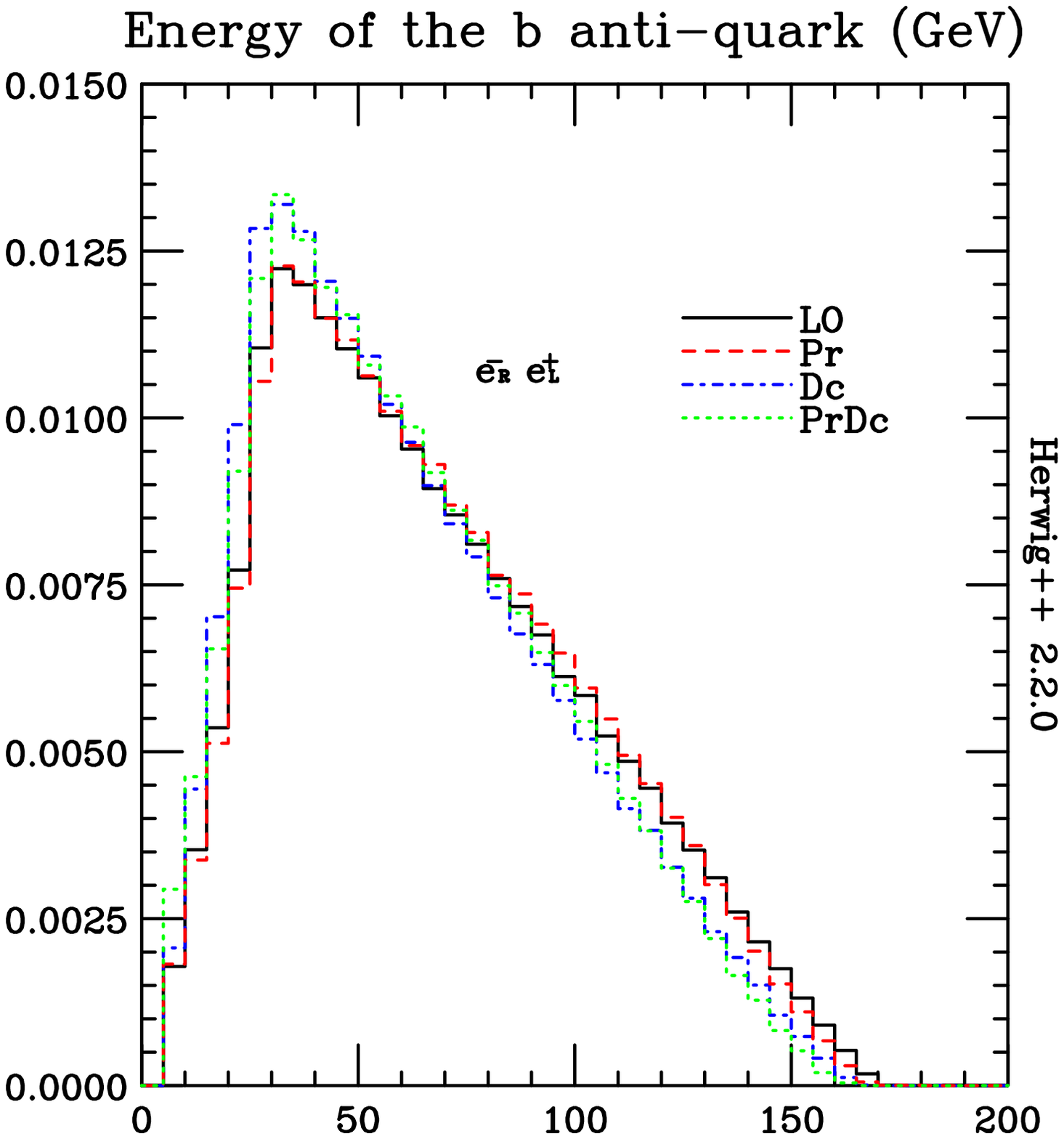,%
width=2.5in,height=3in,angle=90}
\caption{Energy of the b anti-quark before hadronization.}
\label{fig:benergy2}
\end{figure}

\begin{figure}[!ht]
\vspace{2cm}
\hspace{0.5cm}
\psfig{figure=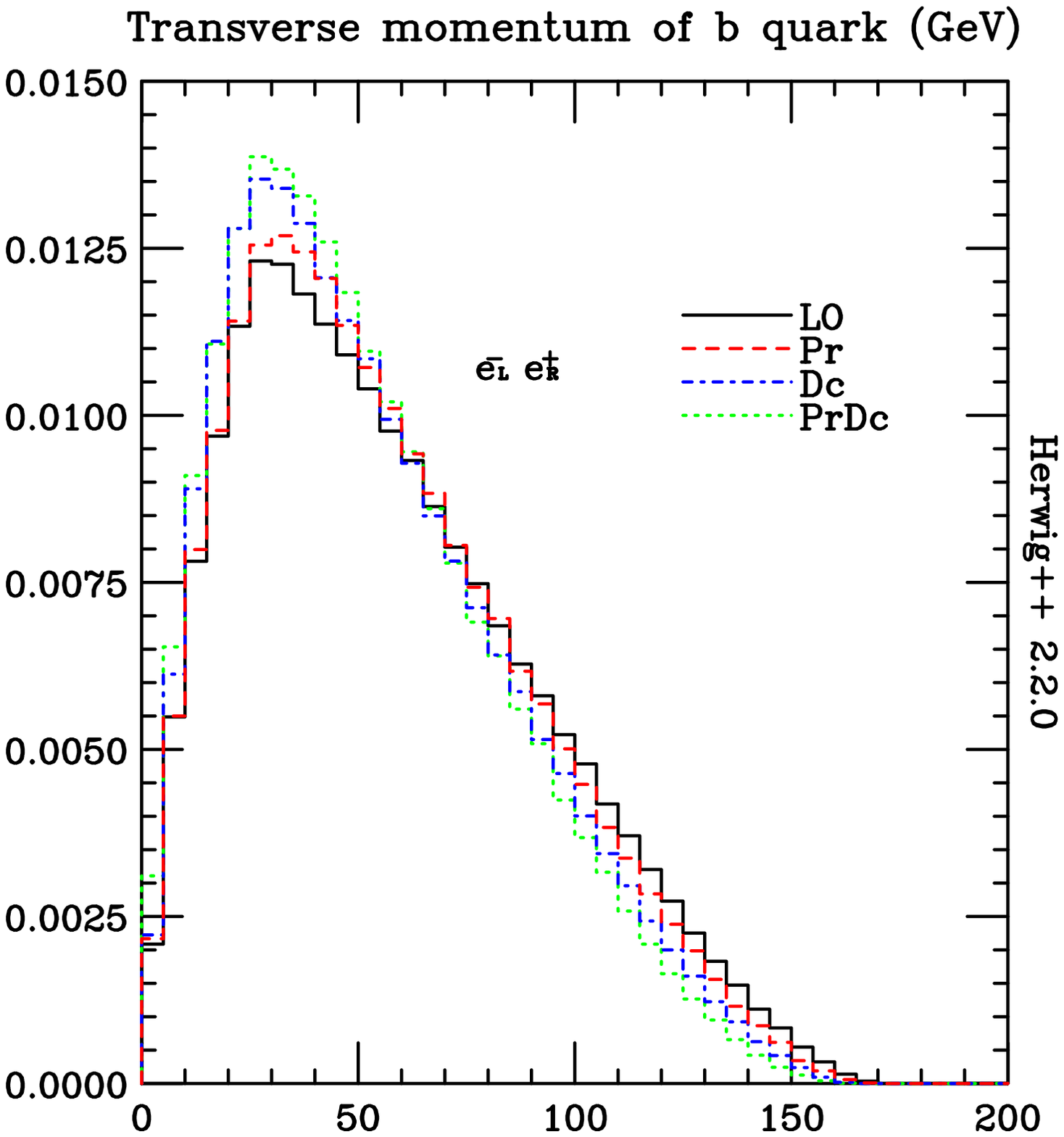,%
width=2.5in,height=3in,angle=90}
\psfig{figure=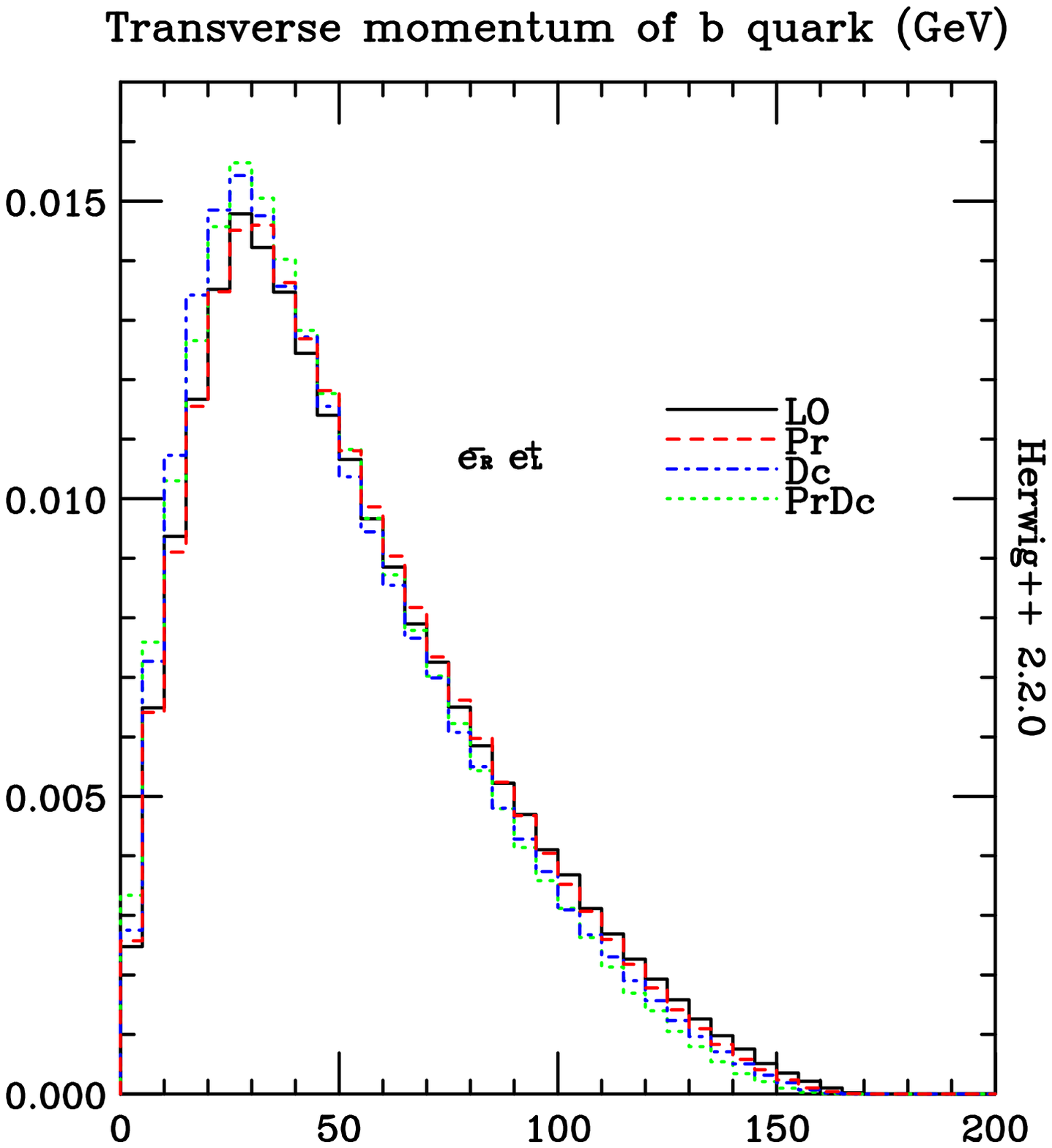,%
width=2.5in,height=3in,angle=90}
\caption{Transverse momentum of the b quark before hadronization.}
\label{fig:bpT}
\end{figure}

\begin{figure}[!ht]
\vspace{2cm}
\hspace{0.5cm}
\psfig{figure=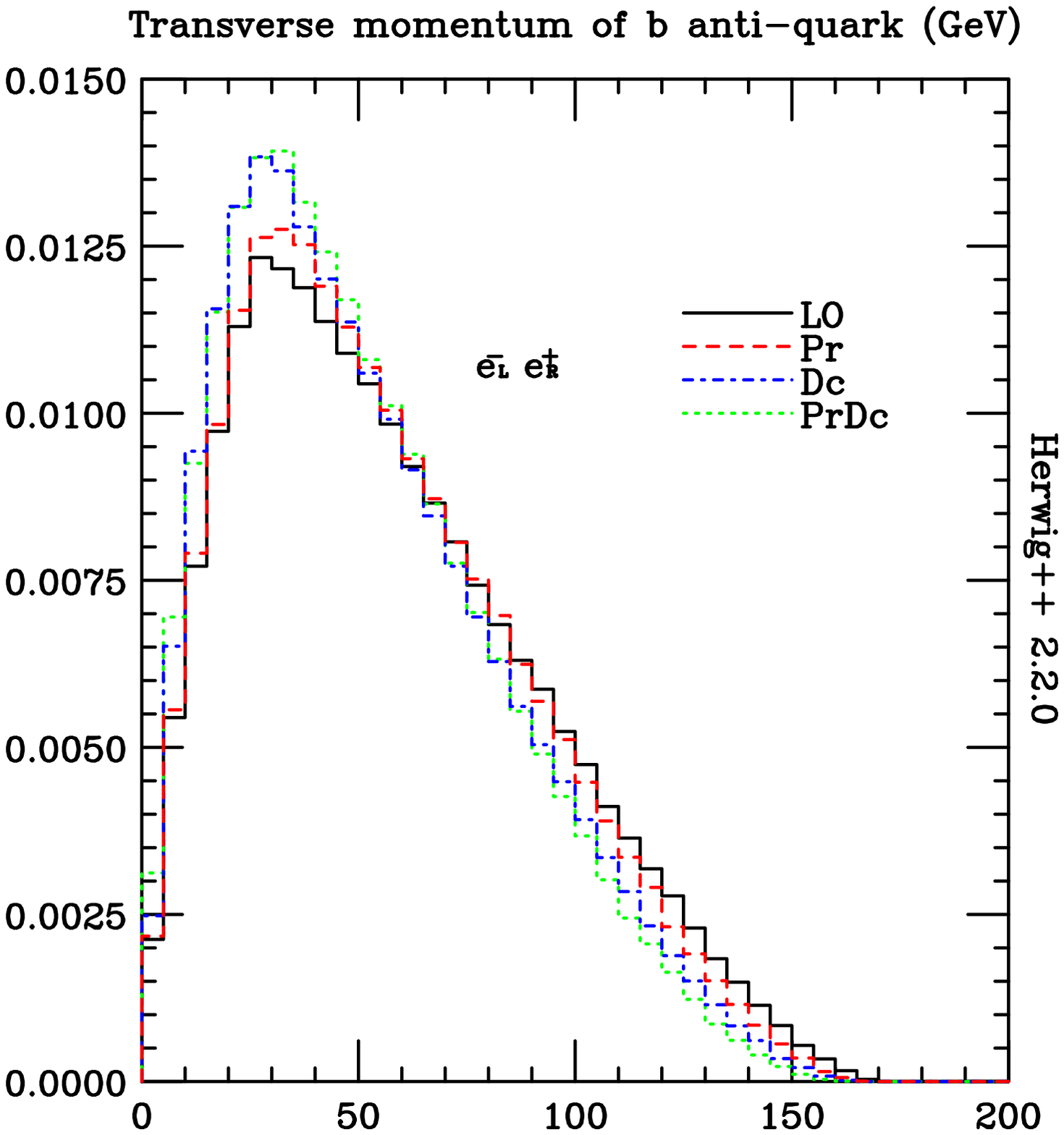,%
width=2.5in,height=3in,angle=90}
\psfig{figure=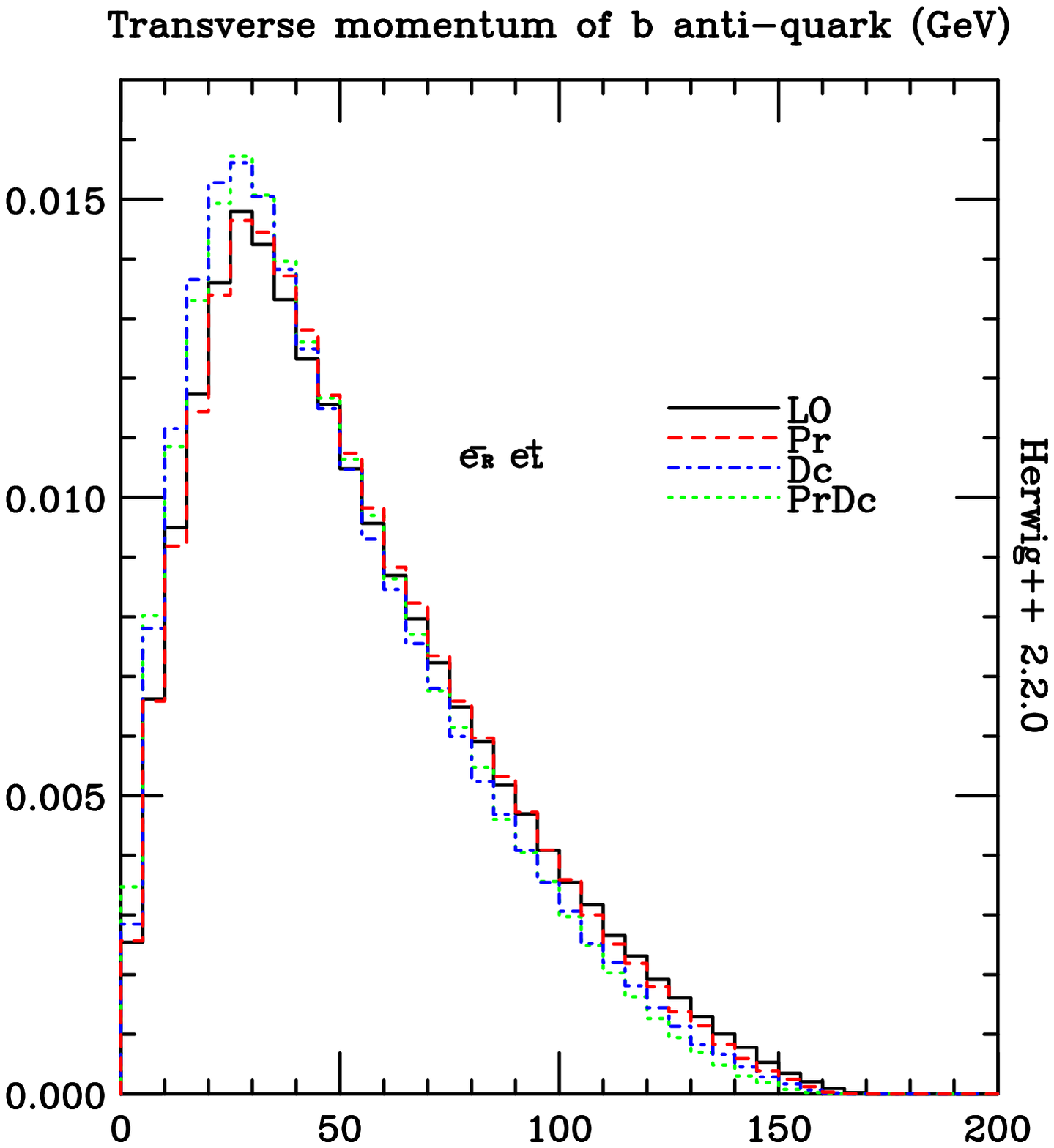,%
width=2.5in,height=3in,angle=90}
\caption{Transverse momentum of the b anti-quark before hadronization.}
\label{fig:bpT2}
\end{figure}

\begin{figure}[!ht]
\vspace{2cm}
\hspace{0.5cm}
\psfig{figure=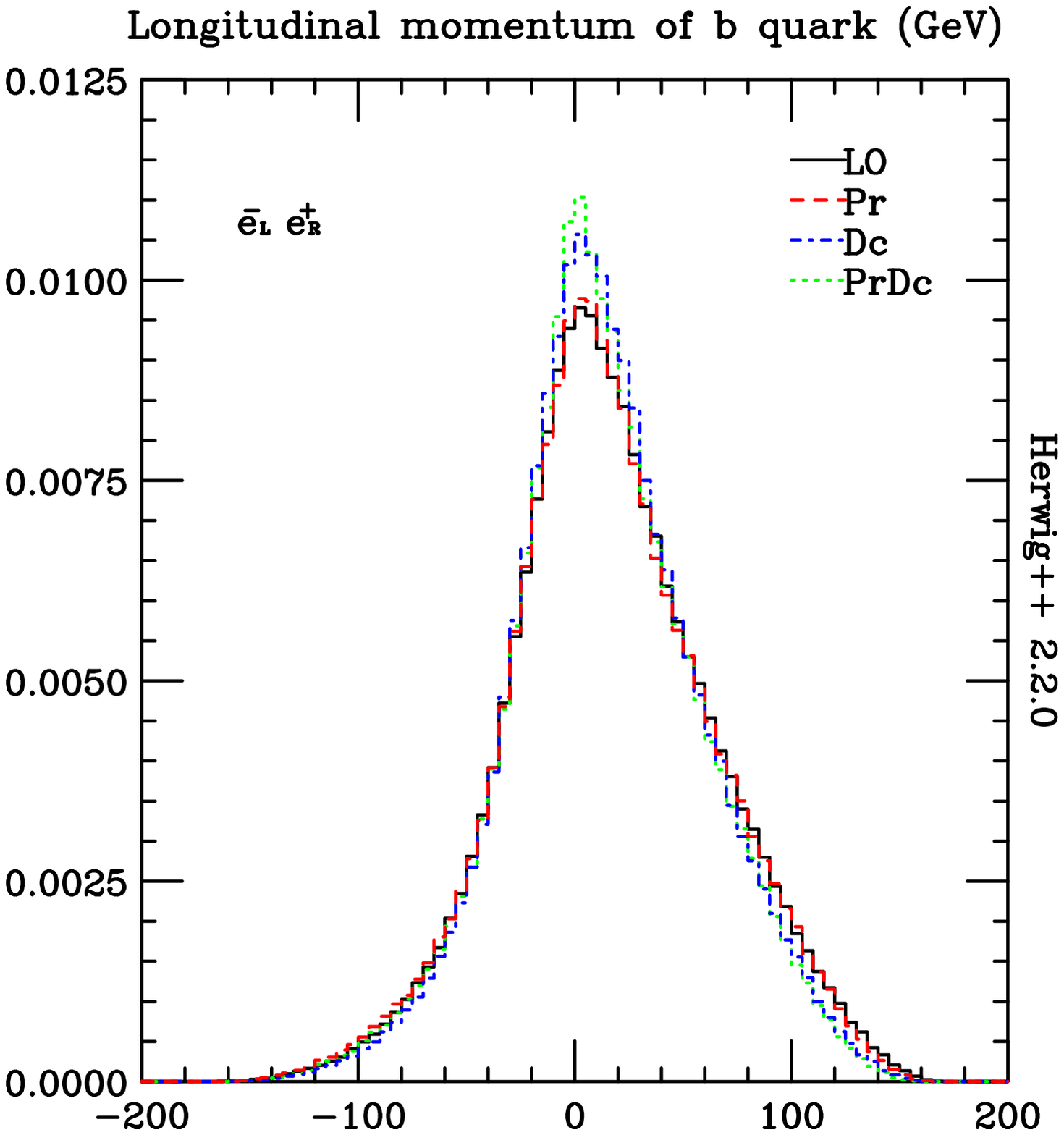,%
width=2.5in,height=3in,angle=90}
\psfig{figure=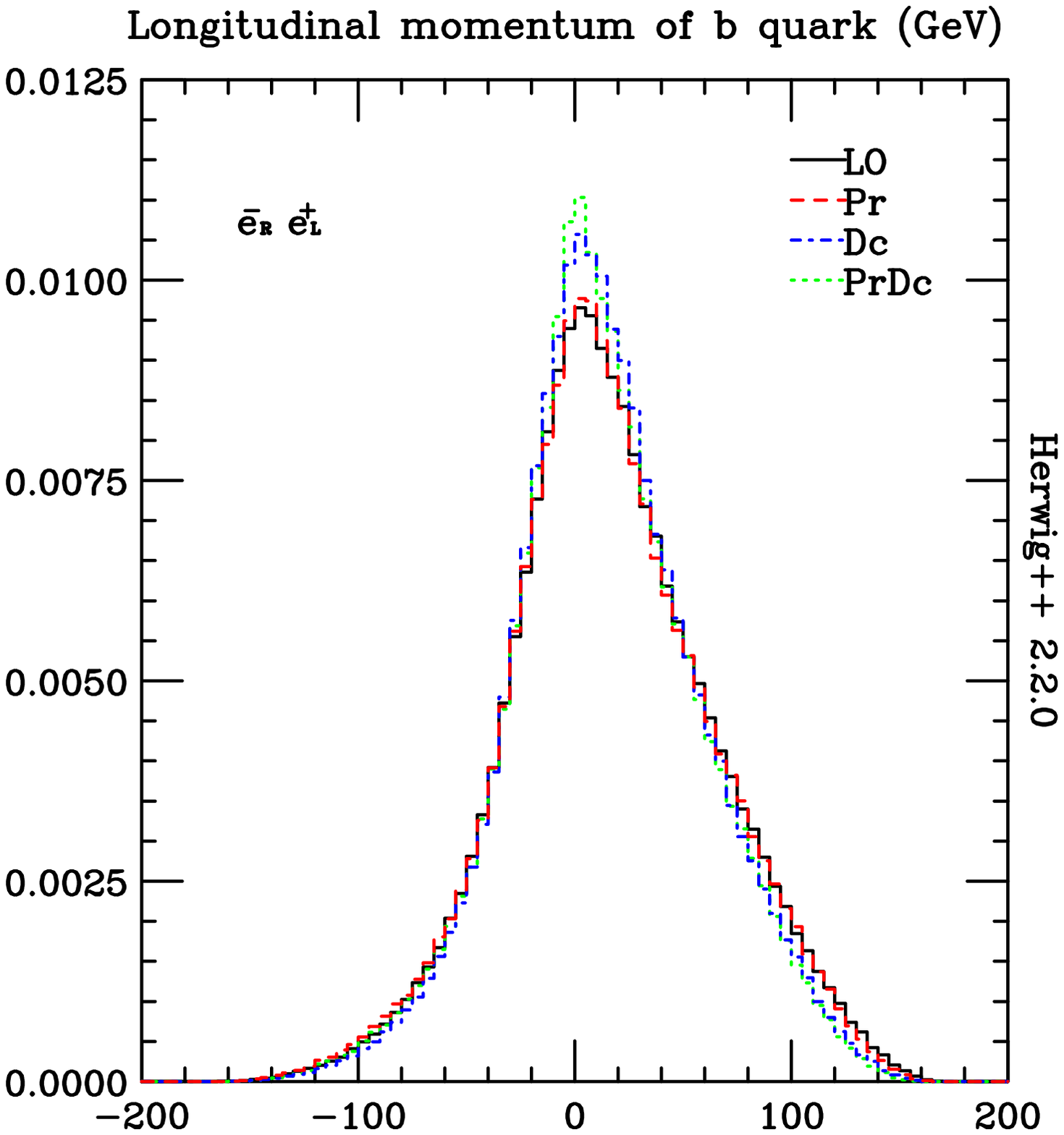,%
width=2.5in,height=3in,angle=90}
\caption{Longitudinal momentum of the b quark before hadronization.}
\label{fig:bpz}
\end{figure}

\begin{figure}[!ht]
\vspace{2cm}
\hspace{0.5cm}
\psfig{figure=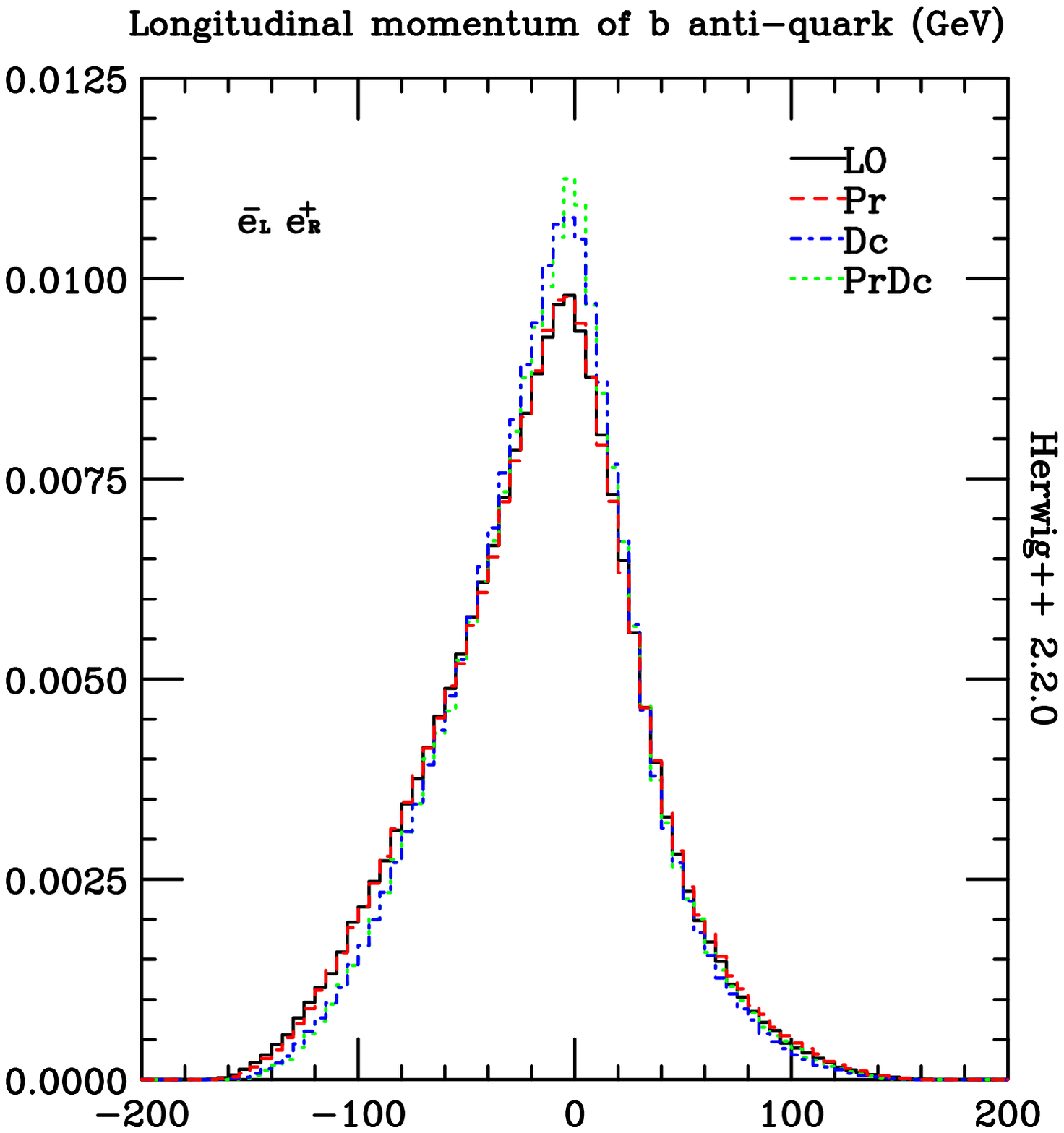,%
width=2.5in,height=3in,angle=90}
\psfig{figure=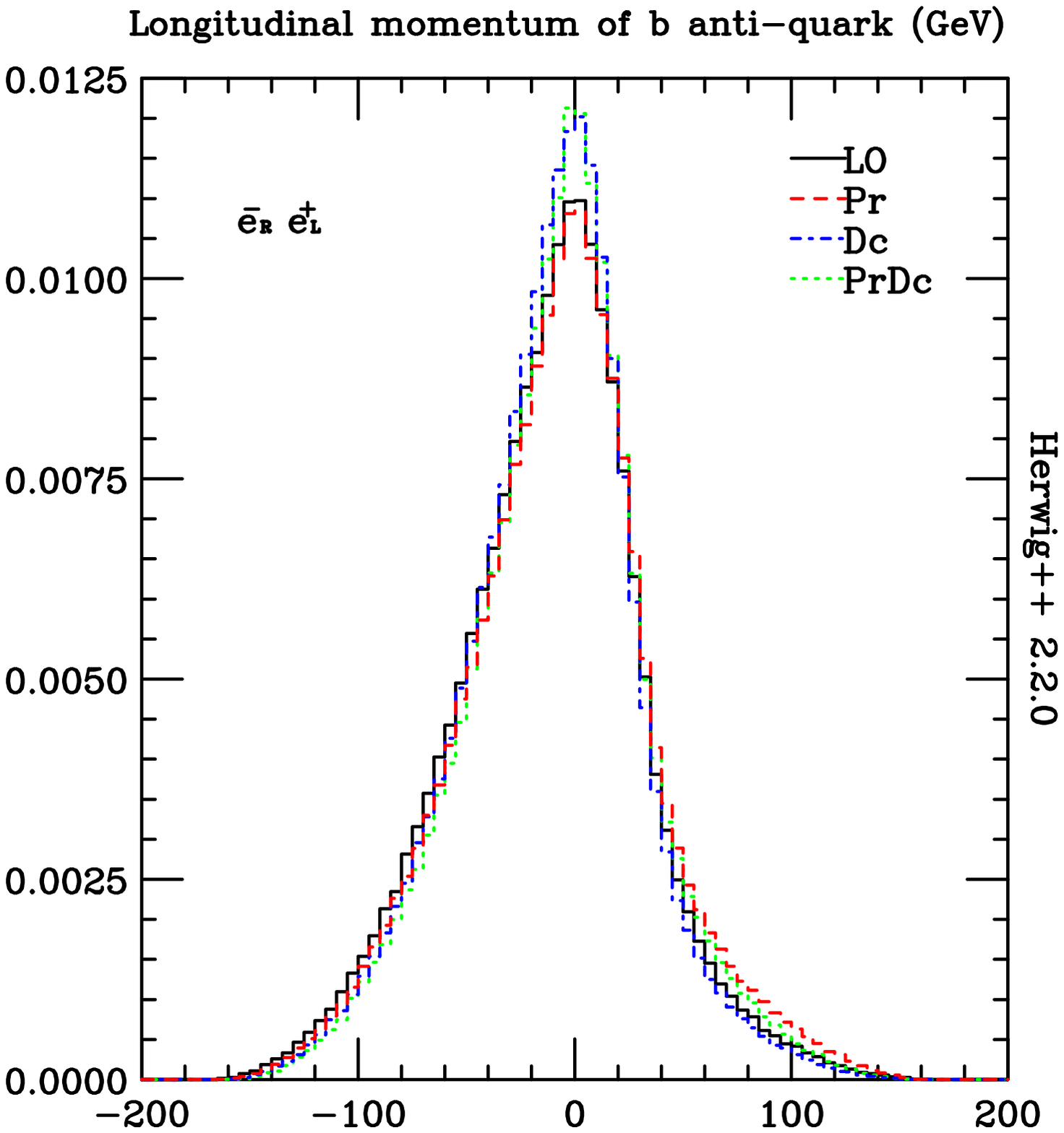,%
width=2.5in,height=3in,angle=90}
\caption{Longitudinal momentum of the b anti-quark before hadronization.}
\label{fig:bpz2}
\end{figure}

Next we interface the generated events with the \textsf{Herwig++ 2.2.0}\cite{Bahr:2008tx} parton shower and veto the hardest emissions
in the production and decay of the top and anti-top pairs. In this section we will consider
collisions at $\sqrt{s}=500$ GeV and only include
the truncated shower for the production emissions. We considered four cases:
\begin{enumerate}
\item
Leading Order (LO): No {\tt POWHEG} emissions.
\item
Production (Pr): Only {\tt POWHEG} emissions in the production are allowed including the
truncated shower.
\item
Decay  (Dc): Only {\tt POWHEG} emissions in the decays of the top/anti-top pairs are allowed.
\item
Production + Decay  (PrDc): Both production and decay emissions are allowed. 
\end{enumerate}
The following distributions
were investigated in the lab frame for the two different $e^+e^-$ initial polarizations:
\begin{enumerate}[i)]
\item
The angle between the lepton from the decay of the top anti-quark and the top quark are
presented in Figure \ref{fig:et}. 
\item
The angle between the lepton and anti-lepton from the decays of the top pairs are
presented in Figure \ref{fig:ee}.
\item
The energy distributions of the $b$ quark and $b$ anti-quark before hadronization are
presented in Figures \ref{fig:benergy} and \ref{fig:benergy2}.
\item
The transverse momenta w.r.t the beam axis of the $b$ quark and $b$ anti-quark before hadronization are
presented in Figures \ref{fig:bpT} and \ref{fig:bpT2}.
\item
The longitudinal momenta (along the beam axis) of the $b$ quark and $b$ anti-quark before
hadronization are presented in Figures \ref{fig:bpz} and \ref{fig:bpz2}.
\end{enumerate}   
\clearpage{}
At leading order, the leptonic correlations in Figures \ref{fig:et} and \ref{fig:ee} are as expected
with higher correlations seen for $e_R^{-}e_L^{+}$ annihilation than for $e_L^{-}e_R^{+}$
annihilation. At next-to-leading order, it can be observed that the
{\tt POWHEG} emissions do not change the shapes of the distributions much except for a
slight broadening of the peaks.

Also at leading order, the distributions in Figures \ref{fig:benergy}-\ref{fig:bpz2} have the expected shapes with the $b$
quarks and anti-quarks having
softer (harder) energy, longitudinal momentum and transverse momentum spectra for
$e_R^{-}e_L^{+}$ ($e_L^{-}e_R^{+}$) annihilation.

Now comparing the {\tt POWHEG} production and decay distributions for the
$b$ quark in Figures \ref{fig:benergy}-\ref{fig:bpz2}, we observe that the decay
emissions soften the spectra more than the production emissions and therefore these have
the greater effect in the production + decay distributions. This is expected since the
scale range available for the production emissions $ \approx \log (\sqrt{s}/m_{\rm t})$ is less than the
range available for the decay emissions $\approx \log (m_{\rm t}/m_{\rm b})$.

\section{Conclusions}
\label{conc}
Using the Monte Carlo event generator \textsf{Herwig++}, we have successfully applied the
{\tt POWHEG} method to investigate angular correlation distributions at next-to-leading
order in top pair production and decays at ILC energies. In all distributions studied, the {\tt POWHEG} emissions have
the effect of broadening the peaks of the leading order predictions slightly. We also
compared momentum distributions of the $b$ quarks and anti-quarks before hadronization and
observe that the decay emissions soften the spectra more at next-to-leading order as expected.  
\section{Acknowledgements}
We are grateful to the other members of the \textsf{Herwig++} collaboration for
developing the program that underlies the present work and for helpful comments. We are particularly grateful to
Bryan Webber for constructive comments and discussions throughout. This research was
supported by the Science and Technology Facilities Council, formerly the Particle Physics
and Astronomy Research Council and the European Union Marie Curie Research Training
Network MCnet.
\bibliography{project}
\bibliographystyle{utphys}
\end{document}